\documentclass[aps,prb,twocolumn,showpacs,amsmath,amssymb,amsthm,superscriptaddress,groupedaddress]{revtex4}
\usepackage{graphics}
\usepackage{epsfig}
\usepackage[colorlinks=true,citecolor=blue,linkcolor=red,linktocpage=true,pagebackref=false]{hyperref}
\usepackage{soul} 
\usepackage[usenames, dvipsnames]{color}
\usepackage{braket}

\begin{document}
\title{
Single-particle emission at finite temperatures 
}
\author{Michael Moskalets}
\email{michael.moskalets@gmail.com}
\affiliation{Department of Metal and Semiconductor Physics, NTU ``Kharkiv Polytechnic Institute", 61002 Kharkiv, Ukraine}

\date\today
\begin{abstract} 
The state of particles injected onto the surface of the Fermi sea depends essentially on the temperature. 
The pure state injected at zero temperature becomes a mixed state if injected at finite temperature. 
Moreover the electron source injecting a single-particle state at zero temperature may excite a multi-particle state if the Fermi sea is at finite temperature. 
Here I unveil a symmetry of the scattering amplitude of a source, which is sufficient to preserve a single-particle emission regime at finite temperatures if such a regime is achieved at zero temperature. 
I give an example and analyze the effect of temperature on time-dependent electrical and heat currents carried by a single-particle excitation. 
\end{abstract}
\maketitle


\section{Introduction}

Often the theoretical models are formulated at zero temperature while the experiment  is always carried out at finite temperatures. 
Nevertheless sometimes a zero-temperature model can be used successfully to explain what is going on in experiment. 
Here I bring an example of how the symmetry of a system promotes a zero-temperature model to work at finite temperatures. 
Namely, I show that a certain symmetry of an electron source protects a single-particle emission regime against a possible disruptive impact of  increasing temperature.   

Implementation of high-frequency on-demand single-electron sources is a major step towards solid-state quantum information processing with fermions \cite{Bennett:2000kl,{Loss:1998ia},Yamamoto:2012bp}. 
It has also a more practical impact, for example, it is promising for realization of the SI unit ampere.\cite{Giblin:2012cl,Pekola:2013fg,{Stein:2015fw}}

One approach to achieve a single-electron emission is to use a very small dot with a quantized electron spectrum. 
The examples are a quantum capacitor \cite{Buttiker:1993wh,Gabelli:2006eg}  used in Ref.~\onlinecite{Feve:2007jx}, a dynamical quantum dot used in Refs.~\onlinecite{Blumenthal:2007ho,{Kaestner:2008gv}}, or even a single donor atom used in Ref.~\onlinecite{Tettamanzi:2014gx}. 

Alternative approach is to use a voltage pulse of a definite shape to excite a single-charge quantum directly from a metallic contact, as it was suggested in Refs.~\onlinecite{{Levitov:1996ie},Ivanov:1997wz} and first experimentally realized in Ref.~\onlinecite{Dubois:2013ul}. 
The corresponding excitations were named {\it levitons}.  

The single-particle nature of electrons injected on-demand was demonstrated by using several ways. 
First, the regime with a minimal high frequency noise was identified as a single-particle emission regime.\cite{Mahe:2010cp} 
Such an identification was supported by the simple quasi-classical model \cite{Albert:2010co} as well as by the full quantum-mechanical calculations based on the Floquet scattering matrix approach \cite{Parmentier:2012ed}. 
Further, the electronic analogue of the famous Hanbury Brown and Twiss (HBT) effect in optics \cite{HanburyBrown:1956bi} was used to count particles  emitted by the source. 
It was demonstrated that no spurious electron-hole pairs were excited during injection of single electrons \cite{Bocquillon:2012if,{Dubois:2013ul}}. 
In addition, the electronic counterpart of the Hong-Ou-Mandel (HOM) effect for photons \cite{Hong:1987gm} demonstrated an anti-bunching expected for fermionic single-particle excitations \cite{Bocquillon:2013dp,{Dubois:2013ul}}.  

Here I am interested in the sources, that inject single electrons close to the surface of the Fermi sea. 
The examples are the source based on a quantum capacitor driven by a harmonic potential, the source of levitons, to name a few. 
In this case one could expect that the ambient temperature would affect significantly the properties of the state emitted by a single-particle source, since the energy of injected particles is comparable with the energy of thermal excitations of the  Fermi sea. 

Indeed, for a quantum capacitor the experiment demonstrates that  the HBT shot noise depends on temperature, it deceases with increasing temperature.\cite{Bocquillon:2012if} 
This effect was attributed to fermionic antibunching of injected particles  and thermal excitations present in the Fermi sea at finite temperatures.\cite{Bocquillon:2012if,{Bocquillon:2013fp},{Marguerite:2016ur}}
The same effect also takes place for levitons.\cite{Dubois:2013fs} 

An alternative explanation was put forward in Ref.~\onlinecite{Moskalets:2015ub},  where the HBT shot noise reduction was related to the fact that the state emitted by the source of levitons at finite temperatures is a mixed state while the one emitted at zero temperature is a pure state. 
The mixed state is less noisy compared to a pure state. 

The outcome of Ref.~\onlinecite{Moskalets:2015ub} is that the state emitted by a single-particle source remains quantum coherent but it is modified by a finite ambient temperature. 
For instance, two fermionic states emitted at finite temperatures preserve their  ability for perfect antibunching during collision (the electronic analogue of the  HOM effect in optics). 
A non-damaging role of temperature (from the point of view of a quantum coherence) is also  demonstrated by the fact that for single-charged levitons the entire shape of the electronic HOM-like signal as a function of the  time delay between two colliding particles is not modified by temperature anymore (though the magnitude decreases with increasing temperature)\cite{{Dubois:2013fs},Glattli:2016tr,{Glattli:2016wl}} even for a random injection \cite{Glattli:2017vp}. 

In the present paper I generalize the analysis carried out in Ref.~\onlinecite{Moskalets:2015ub} to a wider class of electron sources, namely to those whose scattering amplitudes possess the time-energy translation  symmetry. 
I show that such a source emitting single particles at zero ambient temperature works as a single-particle source at finite temperatures as well. 
The effect of finite temperatures is reduced to the fact that a pure state emitted at zero temperature is turned into a mixed state emitted at finite temperatures. 
 
Note that electrons injected from a dynamical quantum dot\cite{Kaestner:2015wu} have energy far above the Fermi energy\cite{Fletcher:2012te}.  
Therefore, one can expect that finite temperatures have no direct effect on the  wave function\cite{Ryu:2016kb,{Kashcheyevs:2017vc}} of injected electrons.  

The paper is organized as follows.
In Sec.~\ref{ecf} the excess correlation function of particles emitted by a single-electron source is introduced. 
Then using the Floquet scattering matrix approach \cite{Moskalets:2011wz} I show how a correlation function at finite temperature can be related to a correlation function at zero temperature. 
Given such a relation one can see that a pure single-particle state  emitted at zero temperature becomes a mixed state if emitted at finite temperature.  
The time-energy translation symmetry of a scattering amplitude of a source is discussed    in Sec.~\ref{tets}.   
A few examples of a source, that preserves a single-particle emission at finite temperatures, are presented in Sec.~\ref{ex}.  
The effect of temperature on time-resolved electrical and heat currents is discussed in Sec.~\ref{temp}. 
I conclude in Sec.~\ref{concl}. 
The Appendix~\ref{appA} contains some auxiliary calculations. 
In particular, the scattering amplitude of a quantum level raising at a constant rapidity calculated in Ref.~\onlinecite{Keeling:2008ft} is derived from the scattering amplitude of a quantum capacitor calculated in Ref.~\onlinecite{Moskalets:2008fz}.

\section{Excess correlation function}
\label{ecf}

The state of a system of non-interacting electrons is fully characterized by  the first-order correlation function ${\cal G}^{(1)}(1;2) = \left\langle \hat\Psi ^{\dag}(1) \hat\Psi(2) \right\rangle$, where $ \hat\Psi(j)$ is a single-particle electron field operator in second quantization evaluated at space-time point $j = 1,2$. 
In a one-dimensional case of interest here the point $j$ is characterized by its coordinate $ x_{j}$ and time $t _{j}$, so $\hat\Psi_{ } (j) \equiv \hat\Psi_{ 
}\left(x_{j}t_{j} \right)$. 
The system I have in mind consists of a periodically driven quantum system, a single-electron source, which is connected to a chiral electron waveguide.
The waveguide in turn is connected to an electron reservoir, a metallic contact. 
The examples of chiral electron waveguides are the edge states in quantum Hall conductors or in topological insulators.\cite{Buttiker:2009bg} 
The electrons in a metallic contact are supposed to be in equilibrium. 
The quantum statistical average $\left\langle \cdots  \right\rangle$ is performed  over the equilibrium state of electrons in the contact they are coming from. 

A working source disturbs an electron system in a waveguide. 
To characterize 
this disturbance 
it is convenient to introduce {\it the excess first-order correlation function}\cite{Grenier:2011js,{Grenier:2011dv},{Haack:2013ch}}, which is defined as the difference of the correlation functions evaluated with the source on and off,

\begin{eqnarray}
G^{(1)}_{ }( t_{1};t_{2}) = {\cal G}^{(1)}_{ on}( t_{1};t_{2}) - {\cal G}^{(1)}_{off }( t_{1};t_{2}) ,
\label{ecf00}
\end{eqnarray}
\ \\ \noindent
where the subscript denotes the status of the source. 
Since all the correlation functions are evaluated at the position of the source (just behind it down the electron stream), we keep only time arguments. 

It is convenient to consider the excess correlation function for electrons in the waveguide as the correlation function for particles  injected by the source into  the otherwise unperturbed waveguide.
Below I will adopt this terminology.  

The excess correlation function $ G^{(1)}$ can be expressed in terms of the  scattering amplitude of the source $S_{in}(t,E)$, which is a quantum mechanical amplitude for an electron with energy $E$ in a waveguide to pass by the source at time $t$.\cite{Moskalets:2008ii} 
While passing by the source an electron can enter the source, stay there during some time, and then come back to the waveguide. 
During its stay within the source an electron is subject to a time-dependent force driving the source. 
This is a reason why the scattering amplitude $S_{in}$ depends on time. 

The relation between the excess correlation function and the scattering amplitude is the following,\cite{Moskalets:2015ub}

\begin{eqnarray}
G^{(1)}_{ }( t_{1};t_{2}) &=&  
\frac{ 1 }{ h v_{ \mu} }
\int   d E    f\left( E \right)  
e^{ \frac{ i }{ \hbar  } E   \left( t_{1} - t_{2} \right)  } 
\nonumber \\
\label{ecf01} \\
&& \times
\left\{
S_{in}^{*}( t_{1}, E) 
S_{in}( t_{2 }, E)    
- 1 \right\} .
\nonumber 
\end{eqnarray}
\ \\ \noindent
Here $f(E) = \left( 1 + e ^{ \frac{ E - \mu }{ k _{B} \theta }}  \right) ^{-1}$ is the Fermi distribution function with  
$ \mu$ and $ \theta$ being the Fermi energy and the temperature, respectively, for electrons in a reservoir the waveguide is connected to;  
$k _{B}$ is the Boltzmann constant. 
Note that an asterisk in $S _{in} ^{*}(t,E)$ denotes the complex conjugation.

\subsection{Zero temperature}

At zero temperature the Fermi distribution function is the step function, $f(E) = \theta\left( E - \mu  \right)$, which is zero at $E > \mu$ and one at $E < \mu$. 
For convenience I introduce a new variable $ \epsilon = E - \mu$ and denote the correlation function at zero temperature by the subscript $0$. 
With this notation we have,

\begin{eqnarray}
G^{(1)}_{ 0 }( t_{1};t_{2}) &=&  
\frac{ e^{\frac{ i }{ \hbar  } \mu   \left( t_{1} - t_{2} \right)   } 
 }{ h v_{ \mu} }
\int\limits_{- \infty}^{0}   d \epsilon 
e^{ \frac{ i }{ \hbar  } \epsilon   \left( t_{1} - t_{2} \right) } 
\nonumber \\
\label{ecf02} \\
&& \times
\left\{
S_{in}^{*}( t_{1}, \epsilon) 
S_{in}( t_{2 }, \epsilon)    
- 1 \right\} ,
\nonumber 
\end{eqnarray}
\ \\ \noindent
where I introduced a short notation $S_{in}(t, \epsilon)  \equiv S_{in}(t, \epsilon + \mu) = S_{in}(t,E) $. 

The equation above can be used as the starting point for the analysis of the state of particles injected by an electron source on the top of the Fermi sea at zero temperature. 

For instance, if the excess correlation function can be represented as the product of two factors dependent on a single time each,

\begin{eqnarray}
G^{(1)}_{0}\left(  t_{1}; t_{2} \right) &=& 
 \Psi ^{*}\left(  t _{1} \right)  \Psi\left(  t _{2}\right) ,
\nonumber \\
\label{ecf03} \\
 \Psi\left(  t \right) &=& e^{ - \frac{i }{ \hbar }  \mu t  } \psi\left( t \right)
\nonumber 
\end{eqnarray}
\noindent \\
then the disturbance of the Fermi sea produced by a working source looks very like as if an electron source would emit a single particle state with wave function $ \Psi\left(  t \right)$ and no a multi-particle state is excited.\cite{Grenier:2013gg,{Moskalets:2015vr}}.
 
This interpretation goes in line with  expectation based on a picture, where an occupied quantum level of the source raises above the Fermi level of electrons in a waveguide, see, e.g.,  Ref.~\onlinecite{Feve:2007jx}.    
 
The specific form of Eq.~(\ref{ecf03}), when it is factorized, is dictated by the symmezry of the correlation function, $G^{(1)}_{0}\left(  t_{1}; t_{2} \right)  =\left[  G^{(1)}_{0}\left(  t_{2}; t_{1} \right) \right] ^{*} $.

\subsection{Finite temperatures}

At non-zero temperature the Fermi function is not a step function anymore.
Therefore, the energy integral in Eq.~(\ref{ecf01}) runs not to zero as in Eq.~(\ref{ecf02}) but to $+ \infty$. 
Nevertheless one can bring $ G^{(1)}$, Eq.~(\ref{ecf01}), into the form resembling $ G^{(1)}_{0}$. 

For this purpose let us use the following identity,

\begin{eqnarray}
f\left(  \epsilon \right) = \int\limits_{ \epsilon}^{ \infty} d \epsilon^{\prime} \left( - \frac{   \partial f}{  \partial \epsilon^{\prime}  }  \right) ,
\label{ecf04}
\end{eqnarray}
\noindent \\ 
change the order of integration in the double integral, which occurs in Eq.~(\ref{ecf01}), 

\begin{eqnarray}
\int\limits_{- \infty}^{ \infty} d \epsilon 
\int\limits_{ \epsilon}^{ \infty} d \epsilon^{\prime} 
\rightarrow 
\int\limits_{- \infty}^{ \infty} d \epsilon^{\prime} 
\int\limits_{- \infty}^{ \epsilon^{\prime}} d \epsilon ,
\label{ecf05}
\end{eqnarray}
\noindent \\
and finally make a shift $ \epsilon \to \epsilon + \epsilon^{\prime}$. 
As a result Eq.~(\ref{ecf01}) becomes, 

\begin{eqnarray}
G^{(1)}_{ }( t_{1};t_{2}) =  
\frac{ e^{ \frac{ i }{ \hbar } \mu  \left( t_{1} - t_{2} \right)  } 
 }{ h v_{ \mu} }
\int\limits_{- \infty}^{ \infty} d \epsilon^{\prime} 
\left( - \frac{   \partial f}{  \partial \epsilon^{\prime}  }  \right)
e^{\frac{ i }{ \hbar  } \epsilon ^{\prime}   \left( t_{1} - t_{2} \right) } 
\nonumber \\
\label{ecf06} \\
\times
\int\limits_{- \infty}^{ 0 } d \epsilon
e^{ \frac{ i }{ \hbar  } \epsilon   \left( t_{1} - t_{2} \right) } 
\left\{
S_{in}^{*}( t_{1}, \epsilon + \epsilon^{\prime}) 
S_{in}( t_{2 }, \epsilon + \epsilon^{\prime})    
- 1 \right\} .
\nonumber 
\end{eqnarray}
\ \\ \noindent

Now we make a crucial step. 
Let us suppose that the scattering amplitude of the source possesses the following time-energy translation symmetry,

\begin{eqnarray}
S_{in}\left( t, \epsilon +  \delta \epsilon \right) = S_{in}\left( t - \delta \epsilon/c , \epsilon \right) ,
\label{ecf07}
\end{eqnarray}
\ \\ \noindent 
where $c$ is a constant. 
Using this equation and  Eqs.~(\ref{ecf06}) and (\ref{ecf02}) we can express the correlation function at finite temperature, $  G ^{(1)}_{ }$, in terms of the correlation function at zero temperature, $  G ^{(1)}_{0}$, as follows, 

\begin{eqnarray}
G^{(1)}_{  }( t_{1};t_{2}) &=&  
\int\limits_{- \infty}^{ \infty} d \epsilon^{\prime} 
\left( - \frac{   \partial f}{  \partial \epsilon^{\prime}  }  \right)
e^{ \frac{ i }{ \hbar } \epsilon ^{\prime}   \left( t_{1} - t_{2} \right)  } 
\nonumber \\
&& \times
G^{(1)}_{ 0 }\left(  t_{1} -  \frac{ \epsilon^{\prime} }{ c  }; t_{2} - \frac{ \epsilon^{\prime} }{ c  } \right) .
\label{ecf08} 
\end{eqnarray}
\ \\ \noindent
This is the central result of the present work. 

The relation above admits an intuitive interpretation in the case if a single-electron emission takes place at zero temperature. 
Substituting Eq.~(\ref{ecf03}) into Eq.~(\ref{ecf08}) we obtain,
\begin{eqnarray}
G^{(1)}_{  }( t_{1};t_{2}) &=&  
\int\limits_{- \infty}^{ \infty} d \epsilon^{\prime} 
\left( - \frac{   \partial f}{  \partial \epsilon^{\prime}  }  \right) 
\Psi_{ \epsilon^{\prime}} ^{*}(t _{1}) 
\Psi_{ \epsilon^{\prime}}(t _{2}) ,
\nonumber \\
\label{ecf09} \\
\Psi_{ \epsilon^{\prime}}(t) &=& e^{ - \frac{ i }{ \hbar  } \left(  \mu + \epsilon^{\prime}  \right)   t }
\psi\left(  t - \epsilon^{\prime}/c \right) .
\nonumber  
\end{eqnarray}
\ \\ \noindent
The correlation function above describes a mixture of single-particle states with component wave functions $ \Psi_{ \epsilon^{\prime}}(t)$ appearing with probability density $-  \partial f/  \partial \epsilon^{\prime}$, which is apparently properly normalized, $\int _{- \infty}^{ \infty } d \epsilon^{\prime} (-  \partial f/  \partial \epsilon^{\prime}) = 1$. 
Note, the different components, $\Psi_{ \epsilon^{\prime}}(t)$ have the  envelope function of the same shape, $ \psi$, but shifted in time. 
An emitted particle looks like it is blurred in time. 

At zero temperature $-  \partial f/  \partial \epsilon^{\prime} = \delta\left( \epsilon^{\prime} \right)$, where $ \delta( \epsilon^{\prime})$ is the Dirac delta function.  
As a result only the component $ \Psi_{0}(t) = e^{ - \frac{ i }{ \hbar  } \mu   t  } \psi\left( t\right)$, i.e. the one with $ \epsilon^{\prime}=0$, survives. 

At non-zero temperature there are many components $ \Psi_{ \epsilon^{\prime}}$ of a mixed state. 
They can be interpreted as follows.   
Let the wave function $ \Psi _{0}(t)$ describes a particle emitted at time $t = t _{0}$ on the top of the Fermi sea filled up to the energy $ \mu$ (i.e., the Fermi sea is at zero temperature). 
Then the wave function $ \Psi_{ \epsilon^{\prime}}(t)$ can be interpreted as the one, which describes the same particle (the same envelope function) but emitted at another time, $t _{0} ^{\prime}= t _{0} +    \epsilon^{\prime}/c$, and on the top of the Fermi sea filled up to another energy, $ \mu ^{\prime} = \mu+ \epsilon^{\prime}$. 
In fact, we do not need to think $ \mu ^{\prime}$ as the Fermi energy of some effective (or fictitious) Fermi sea. 
Better to say, that $ \mu ^{\prime}$ defines the lower bound of energy for an emitted electron. 
This energy enters the corresponding phase factor in Eq.~(\ref{ecf09}). 
Note that the state of an emitted particle is a superposition of states with different energies,\cite{{Keeling:2006hq},Battista:2013ew,{Battista:2014tj}} which can deviate from the lower band $ \mu ^{\prime}$.   
This deviation is encoded into a time dependence of the envelope function $ \psi(t)$.\cite{Moskalets:2014ea}  

The shift of the emission time for different components,  $t _{0} ^{\prime}= t _{0} +    \epsilon^{\prime}/c$, is consistent with the interpretation of $ \mu ^{\prime} = \mu+ \epsilon^{\prime}$ as the minimal energy. 
To illustrate it let us consider a source consisting of a single occupied quantum level ${\cal E}(t)$, which is elevated up in energy. 
An electron can leave the source when unoccupied states in the Fermi sea outside become available.  

At zero temperature, the edge of the Fermi distribution function is sharp and the unoccupied states become available only when the energy of a quantum level will exceed the Fermi energy $ \mu$. 
Thus, at zero temperature the time of emission $ t _{0}$ is defined by the following equation ${\cal E}(t _{0}) = \mu$.  

In contrast, at non-zero temperature the edge of the Fermi distribution function is smeared out. 
Therefore, an electron can escape from the source even when ${\cal E} < \mu$. 
In addition, an electron can stay in the dot even when ${\cal E} > \mu$, since the partially occupied levels outside hamper its escape. 
Therefore, there are many possibilities to escape, that results in many contributions in Eq.~(\ref{ecf09}). 
Each component describes an emission process, which starts when the energy of a level becomes larger than some definite energy $ \mu^{\prime}$.   
If an electron escapes at $ \mu ^{\prime} \ne \mu$, then the time of emission is defined as follows: $ {\cal E} \left(  t _{0} ^{\prime} \right) = \mu ^{\prime} $, where $ t _{0}^{\prime} \ne t_{0}$. 
The difference $t _{0}^{\prime} - t_{0}$ is exactly the time necessary to change the energy of a level ${\cal E}(t)$ from $ \mu$ to $\mu ^{\prime}$. 
Note, that all the components in question are mutually incoherent -- hence they constitute a mixture state -- since they are well distinguishable by their time of emission. 

Note that in Eq.~(\ref{ecf09}) the difference $t _{0}^{\prime} - t _{0}$ is linear in $ \epsilon ^{\prime} =  \mu^{\prime} - \mu$. 
This property is a direct consequence of the symmetry of the scattering amplitude given in Eq.~(\ref{ecf07}).       

Now we discuss the conditions, which can lead to such a symmetry of the scattering amplitude.

\section{Time-energy translation symmetry of the scattering amplitude}
\label{tets}

Simple analysis shows that Eq.~(\ref{ecf07}) is satisfied if the scattering amplitude depends on a single argument, $ \epsilon - c t$, rather than on time $t$ and energy $ \epsilon$ separately, 

\begin{eqnarray}
S _{in}\left(  t, \epsilon \right) \equiv S _{in} \left(  \epsilon - c t \right) .
\label{te01}
\end{eqnarray}
\noindent \\
In this case the scattering amplitude is invariant under the simultaneous translation in energy by an amount of $ \delta \epsilon$ and in time by an amount of $ \delta \epsilon /c$ in full agreement with Eq.~(\ref{ecf07}). 

To clarify conditions for such invariance to hold let us proceed as follows.
For the source side-attached to a one-dimensional electron waveguide, the scattering amplitude is expressed in terms of a phase $ \varphi(t, \epsilon)$ accumulated by an electron during its stay within the source, $S_{in}(t, \epsilon)  \sim e^{i \varphi(t, \epsilon)}$. 
The source is driven by the time-dependent potential $U(t)$. 

To relate $U(t)$ and $ \varphi(t, \epsilon)$ let us consider a simple model: The source is a one-dimensional ballistic loop of length $L$.
For a ballistic motion inside the source, the phase $ \varphi$  can be effectively represented as follows, $ \varphi(t, \epsilon) \sim k(\epsilon) L - (e/ \hbar) \int _{t - \tau_{D}}^{t } dt^{\prime} U(t^{\prime})$, where $k$ is an electron wavenumber and $ \tau_{D}$ is the dwell time - an effective time, during which an electron stays within the source. 

The next step, we linearize the dispersion relation. 
If the Fermi energy is the largest energy scale in the problem, we can write  $k(\epsilon) \approx k(0) + \epsilon/( \hbar v_{ \mu})$, where $v_{ \mu}$ is a velocity of an electron with Fermi energy. 

And finally, we suppose that the potential $U(t)$ changes not too fast, 
such that 
it can be linearized within the relevant time interval (the time interval of duration $ \tau_{D}$), $eU(t) \approx eU _{0}  + c t$, where  $c = dU/dt $ is the rapidity. 

Within these approximations the phase becomes, $ \varphi(t, \epsilon) \approx \varphi_{0} + \left( \epsilon - c t  \right) ( \tau_{D}/ \hbar)$, where $ \varphi_{0}$ is independent of both $ \epsilon$ and $t$. 
It is clear that the phase $ \varphi$ -- hence the scattering amplitude $S_{in}$ -- is of the form required by Eq.~(\ref{te01}). 
Recall that we used both a linearized energy spectrum and a linearized in time driving potential.  

Below I analyze some models of a single-electron source known from the literature and discuss the temperature effect.

\section{Examples}
\label{ex}

\subsection{Single quantum level raising at a constant rapidity}
\label{c}

In Ref.~\onlinecite{Keeling:2008ft} an electron transfer between a localized state and the Fermi sea at zero temperature was analyzed. 
It was shown that if the energy of a localized state increases linearly in time, ${\cal E} = c t$, then such a transfer is noiseless. 
That is, when a localized state raises at a constant rapidity above the Fermi level, an electron is emitted into the Fermi sea while no additional electron-hole pairs  are excited.  
The scattering amplitude, describing such a process, is \cite{Keeling:2008ft}

\begin{eqnarray}
S_{in} ^{(c)}(t, \epsilon) &=&  
1-  \int\limits _{0}^{ \infty } d \xi 
e^{ - \frac{ \xi }{2 } } 
e^{ \frac{ i }{ \hbar  } \xi  \tau _{D}  \left(  \epsilon - c t \right) } 
e^{  i \xi^{2} \frac{ \zeta }{ 4 }  }
,
\label{ex01}
\end{eqnarray}
\ \\ \noindent
where a parameter $ \zeta = 2c \tau ^{2} _{D}/\hbar$.

Obviously the amplitude $S_{in} ^{(c)}$ is of the form given in Eq.~(\ref{te01}). 
Therefore, one can conclude that, according to Eq.~(\ref{ecf09}),  such a source remains a single-electron source even if the temperature of the Fermi sea is non zero. 

At zero temperature an electron emitted is in a pure state with wave function \cite{Keeling:2008ft}

\begin{eqnarray}
\Psi ^{(c)}\left( t \right) = 
\frac{  e^{ - \frac{ i }{ \hbar  } \mu  t  } }{  \sqrt{\pi  \Gamma _{\tau} v_{ \mu}} } 
\int\limits_{0}^{ \infty} \frac{ d \epsilon  }{ 2\epsilon _{0}  }
e^{ -  \frac{ \epsilon  }{ 2\epsilon _{0}  } } 
e^{  - \frac{ i }{ \hbar  } \epsilon  t   } 
e^{ i \left(   \frac{ \epsilon  }{ 2\epsilon _{0}  } \right) ^{2} \zeta  } ,  
\label{ex02}
\end{eqnarray}
\noindent \\
where $ \epsilon _{0} = \hbar /(2 \Gamma _{\tau})$ is the expectation value of energy of an emitted electron whose extent in time is $ 2 \Gamma _{\tau} = \hbar / (c \tau _{D})$. 

At finite temperatures an electron emitted is in a mixed state, see  Eq.~(\ref{ecf09}), with component wave functions,  

\begin{eqnarray}
\Psi _{ \epsilon ^{\prime}} ^{(c)}\left( t \right) = 
\frac{  e^{ - \frac{ i }{ \hbar  } \left(  \mu + \epsilon ^{\prime} \right)  t  } }{  \sqrt{\pi  \Gamma _{\tau} v_{ \mu}} } 
\int\limits_{0}^{ \infty} \frac{ d \epsilon  }{ 2 \epsilon _{0}  }
e^{ -  \frac{ \epsilon  }{ 2\epsilon _{0}  } } 
e^{  - \frac{ i }{ \hbar  } \epsilon  \left(  t - \frac{ \epsilon ^{\prime} }{c  }\right)   } 
e^{ i \left(   \frac{ \epsilon  }{ 2\epsilon _{0}  } \right) ^{2} \zeta  } .  
\nonumber \\
\label{ex03}
\end{eqnarray}
\noindent \\

\subsection{Quantum capacitor as a single-electron source}

An on-demand single-electron source based on a quantum capacitor\cite{Buttiker:1993wh,Gabelli:2006eg}  was realized experimentally in Ref.~\onlinecite{Feve:2007jx}. 
The quantum capacitor was made of a short circular edge state of a two-dimensional electron gas being in the integer quantum Hall effect regime. 
The nearby metallic gate -- which is used to change position of quantum levels -- screens effectively electron-electron Coulomb interactions. 
This is a reason why the theory of non-interacting electrons has proved useful to describe this source and properties of an emitted electron quantum state.\cite{Moskalets:2008fz,{Moskalets:2013dl}} 

A quantum capacitor is modeled as a one-dimensional ballistic quantum dot (1DQD), which is connected to the Fermi sea via  a quantum point contact (QPC) with reflection/transmission amplitude $r/\tilde{t}$.\cite{Gabelli:2006eg} 
The circumference of a 1DQD is $L$. 
A periodic in time electrical potential $U(t) = U(t + {\cal T})$, applied to a nearby gate, is used to change energy of quantum levels in the dot. 
The scattering amplitude of a 1DQD reads \cite{Moskalets:2008fz}

\begin{eqnarray}
S_{in} ^{(cap)}(t,E) = r + \tilde{t}^2  \sum_{q=1}^{\infty}r^{q-1} 
e^{i\left\{q k(E) L - \Phi_{q}(t)\right\}}, 
\label{ex04}
\end{eqnarray}
\noindent \\ 
where $k(E)$ is a wavenumber of an electron in the Fermi sea, which  enters a 1DQD and, after $q$ revolutions along the circumference of a 1DQD, accumulates a time-dependent phase 

\begin{eqnarray}
\Phi_{q}(t)=\frac{e}{\hbar}\int_{t-q\tau(E)}^{t}dt^{\prime}U(t^{\prime}) . 
\label{ex05}
\end{eqnarray}
\ \\ \noindent
Here $ \tau(E) = L/ v(E)$ is the time of a single revolution, $v(E)$ is an electron velocity. 

Notice, each term in Eq.~(\ref{ex04}) describes a partial process, when an electron with energy $E$ enters a dot, makes $q$ revolutions, and leaves a dot at time $t$.\cite{Moskalets:2008ii}
The first term corresponds to an electron reflected back to the waveguide  without entering the dot. 
Note that the dwell time for this model is defined as $ \tau _{D} = \tau / T$, where $T = \left | \tilde t \right | ^{2}$ is the transmission probability of a QPC connecting the dot and the waveguide.  

Below we discuss several protocols of drive $U(t)$, that guarantees a single-particle emission.

\subsubsection{A harmonic potential}

In Appendix~\ref{appA} we show that in the case case of a harmonic drive, $U(t) = U _{0} \cos \left(  \Omega t \right)$,  and under the following conditions,

\begin{eqnarray}
\mu  \gg   eU _{0} \sim \Delta \gg \hbar \Omega, \delta ,
\label{ex06} 
\end{eqnarray}
\noindent \\ 
($ \mu$ is the Fermi energy, $ \Delta$ is the level spacing, $ \delta$ is the level width) the scattering amplitude $S _{in}^{(cap)}$, Eq.~(\ref{ex04}), can be cast into the form of amplitude  $S _{in} ^{(c)}$, Eq.~(\ref{ex01}).
This fact allows us to use the results of Sec.~\ref{c} to describe the state emitted by a harmonically driven quantum capacitor. 
In particular, the state emitted by a quantum capacitor at zero temperature is a pure single-particle state with wave function $ \Psi ^{(c)}$, Eq.~(\ref{ex02}). 
At not very high temperatures, $k _{B} \theta \ll \Delta$, the emitted state remains a single-particle state but becomes a mixed state with correlation function given in Eq.~(\ref{ecf09}) and component wave functions given in Eq.~(\ref{ex03}). 

Let us briefly comment on the inequalities given above. 
The inequality $ \mu \gg eU _{0} \sim \Delta$, guaranties that we can linearize the dispersion relation for electrons injected from the source into the waveguide. 
The next inequality,  $ \Delta \gg \delta, k _{B} \theta$, ensures that only a single level is involved into the process of emission.   
In addition, since $ eU _{0} \gg \delta$ then throughout the emission process the rapidity of a quantum level, $c = edU/dt$, can be regarded as constant. 
Note that the time characteristic for emission is {\it the crossing time}, during which the quantum level of width $ 2 \delta$ crosses the Fermi level.  

The rapidity $c$ is useful to define working regimes of a source, adiabatic and non-adiabatic. 
Roughly speaking, the adiabatic regime is realized at vanishingly small rapidity, $c \to 0$, while the non-adiabatic regime is realized at finite rapidity. 
To be more specific, with increasing rapidity the crossover from adiabatic to non-adiabatic regime occurs when the dwell time $ \tau_{D} = \hbar /(2 \delta)$ becomes of the order of the crossing time  $  2\Gamma _{\tau} = 2 \delta / c$.\cite{Splettstoesser:2008gc}

A harmonically driven quantum capacitor in the adiabatic emission regime at zero temperature was analyzed in Refs.~\onlinecite{{Haack:2013ch},{Moskalets:2013dl}}. 
It was found there that the wave function of an electron emitted adiabatically is,  

\begin{eqnarray}
\Psi ^{(ad)}\left(  t \right) = \frac{ e ^{- \frac{ i }{ \hbar  } \mu t } }{ \sqrt{ \pi \Gamma _{\tau} v _{ \mu}}} \frac{1}{t/ \Gamma _{\tau} - i } .
\label{ex07}
\end{eqnarray}
\noindent \\
This wave function (up to an irrelevant phase factor) is a limit of $ \Psi ^{(c)}$, Eq.~(\ref{ex02}),  at $ \zeta = \tau_{D}/  \Gamma _{\tau} \to 0$. 
Therefore, the parameter $ \zeta = 2c \tau _{D} ^{2}/ \hbar = c \hbar /\left( 2 \delta ^{2}  \right)$ in  Eq.~(\ref{ex02}) can be considered as {\it the adiabaticity parameter}. 

Importantly, the inequalities (\ref{ex06}) do not put any restrictions on the adiabaticity parameter $ \zeta$.  
Moreover, if these inequalities are satisfied then even at arbitrary, not necessarily harmonic, drive a quantum capacitor can work as a single-particle source. 
Note that at arbitrary drive the frequency related to a periodicity, $ \Omega = 2 \pi / {\cal T}$, should be replaced by the frequency $ \Omega ^{\prime}$, which characterizes how fast a driving potential changes during the crossing stage.

\subsection{The source of levitons}

Not only the energy-time symmetry of a scattering amplitude, which is expressed in Eq.~(\ref{ecf07}), enables to relate finite-temperature and  zero-temperature correlation functions of the state emitted by an electron source. 
Another possibility arises in the case of energy-independent scattering amplitude. 

For example, a metallic contact with electrical potential different from that of the other contacts can be effectively described by an energy-independent scattering amplitude.  
Let us denote the corresponding potential as $V(t)$ (in the case of a DC bias, $V$ is independent of time $t$). 
The effective scattering amplitude reads, $S _{in} ^{(V)}(t) = \exp\left( - \frac{i e}{ \hbar}  \int _{ }^{ t } dt ^{\prime} V(t ^{\prime})  \right)$. 
Using this amplitude we can easily relate a finite-temperature correlation function $G ^{(1)}$, Eq.~(\ref{ecf06}), and the correlation function at zero temperature $G ^{(1)} _{0}$, Eq.~(\ref{ecf02}), as follows,\cite{Moskalets:2015ub}

\begin{eqnarray}
G^{(1)}_{  }( t_{1};t_{2}) &=&  
\int\limits_{- \infty}^{ \infty} d \epsilon^{\prime} 
\left( - \frac{   \partial f}{  \partial \epsilon^{\prime}  }  \right)
e^{ \frac{ i }{ \hbar } \epsilon ^{\prime}   \left( t_{1} - t_{2} \right)  } 
G^{(1)}_{ 0 }( t_{1}; t_{2}) .
\nonumber \\
\label{ex09} 
\end{eqnarray}
This relation is of the form of Eq.~(\ref{ecf08}), where we formally set $c \to \infty$. 

With some specific choice of a time-dependent potential $V(t)$ the metallic contact can serve as a single-electron source. 
As it was predicted theoretically long ago \cite{Levitov:1996ie,Ivanov:1997wz} and recently was confirmed experimentally \cite{Dubois:2013ul,Jullien:2014ii}, at zero temperature the Lorentzian in shape voltage pulse $eV(t) =  n^{\star} \frac{2 \hbar \Gamma _{\tau}}{t ^{2} + \Gamma _{\tau} ^{2} }$ (with integer $  n^{\star} = 1, 2, \dots$) excites a particle carrying an integer charge $q = e  n^{\star}$ with the Fermi sea remaining intact (no electron-hole pairs are excited). 
An excitation with $  n^{\star} = 1$ was named {\it a leviton}.\cite{Dubois:2013ul}  
The wave function of a leviton, $ \Psi^{(L)}(t)$, is the one given in Eq.~(\ref{ex07}).\cite{Keeling:2006hq,Dubois:2013fs} 

As it is clear from Eq.~(\ref{ex09}), the source of levitons remans a single-particle source even at finite temperatures. 
The only new ingredient arising at finite temperatures is that the state of a leviton becomes a mixed state with the following component wave functions,\cite{Moskalets:2015ub}

\begin{eqnarray}
\Psi _{ \epsilon ^{\prime} } ^{(L)}(t) = e ^{ - \frac{i}{ \hbar} \left( \mu + \epsilon ^{\prime} \right) t } \psi ^{(L)} (t)  .
\label{ex10}
\end{eqnarray}
\noindent \\
Here $  \psi ^{(L)} (t) = \frac{ 1 }{ \sqrt{ \pi \Gamma _{\tau} v _{ \mu}}} \frac{1}{t/ \Gamma _{\tau} - i } $ is the envelope wave function.
Notice the absence of a time shift in the argument of the envelope  function in Eq.~(\ref{ex10}). 
This is in contrast with  Eq.~(\ref{ecf09}), where the shift in time in the envelope function for a particular component $\Psi_{ \epsilon^{\prime}}$ is inversely proportional to the rapidity, $ \sim \epsilon ^{\prime}/c$.

\section{Temperature effect  on single-electron transport characteristics}
\label{temp}

As we showed above, the finite temperatures (in many cases) does not ruin a single-particle emission regime. 
Therefore, the fact, that in real experiment the temperature is not zero, does not compromise a single-particle source emitting particle even close to the surface of the Fermi sea. 
A remarkable property of such sources is that they put an electron onto the surface of the Fermi sea in a very gentle way, in a way, which allows to create states, which can demonstrate subtle quantum effects. 

An example is a state emitted by a harmonically driven quantum capacitor. 
As we already mentioned, a corresponding scattering amplitude $S _{in} ^{(cap)}$, Eq.~(\ref{ex04}), is the sum of partial amplitudes corresponding to different time intervals (of duration $q \tau$). 
Interference of these amplitudes,  can cause oscillations of physical quantities, such as, for example, single-particle electrical and heat currents \cite{Moskalets:2016va}. 
Therefore, one can say that these oscillations are manifestation of {\it the interference in time}. 
Close related phenomenon is the diffraction in time . \cite{Moshinsky:1952iz,delCampo:2009wn} 

Below we investigate how finite temperatures -- unavoidably present in experiment -- modify these oscillations.

\subsection{Time-dependent electrical current}

An electrical current is expressed in terms of the correlation function as follows (see, e.g. Ref.~\onlinecite{Moskalets:2016va}), 
\begin{eqnarray}
I_{ }(t) = e v_{ \mu} G^{(1)}_{ }( t; t) . 
\label{e01}
\end{eqnarray}
\ \\ \noindent
Equation (\ref{ecf08}) allows us to relate a current at finite temperatures, $I _{ \theta}$, to a zero-temperature current $I _{0}(t) = e v _{ \mu} G _{0} ^{(1)}(t;t)$ as follows,  
\begin{eqnarray}
I _{ \theta}(t) = 
\int\limits_{- \infty}^{ \infty}    d \epsilon ^{\prime}  
\left( - \frac{  \partial f( \epsilon ^{\prime}) }{  \partial \epsilon ^{\prime} } \right) 
I_{0}\left ( t - \frac{ \epsilon ^{\prime} }{ c } \right) .
\label{e02}
\end{eqnarray}

Note that in the case of a leviton -- the corresponding correlation function is given in Eq.~(\ref{ex09}) -- the current is not affected by finite temperatures at all, $I _{ \theta} ^{(L)}(t) = I _{0} ^{(L)}(t)$. 
This fact emphasizes an essential difference between a leviton and an electron emitted by a quantum capacitor working in the adiabatic regime [though the envelope function looks the same, please, compare Eqs.~(\ref{ex07}) and (\ref{ex10})].  
Such a difference also manifests itself in a high-frequency noise, present in the case of a quantum capacitor\cite{Moskalets:2013kj} but absent in the  case of the source of levitons.

\subsubsection{Single quantum level raising at a constant rapidity}

Let us analyze a current carrying by an electron emitted onto the surface of the Fermi sea from a quantum level, whose energy increases at a constant rapidity, ${\cal E}  = c t$. 

\begin{figure*}[t]
\includegraphics[width=80mm]{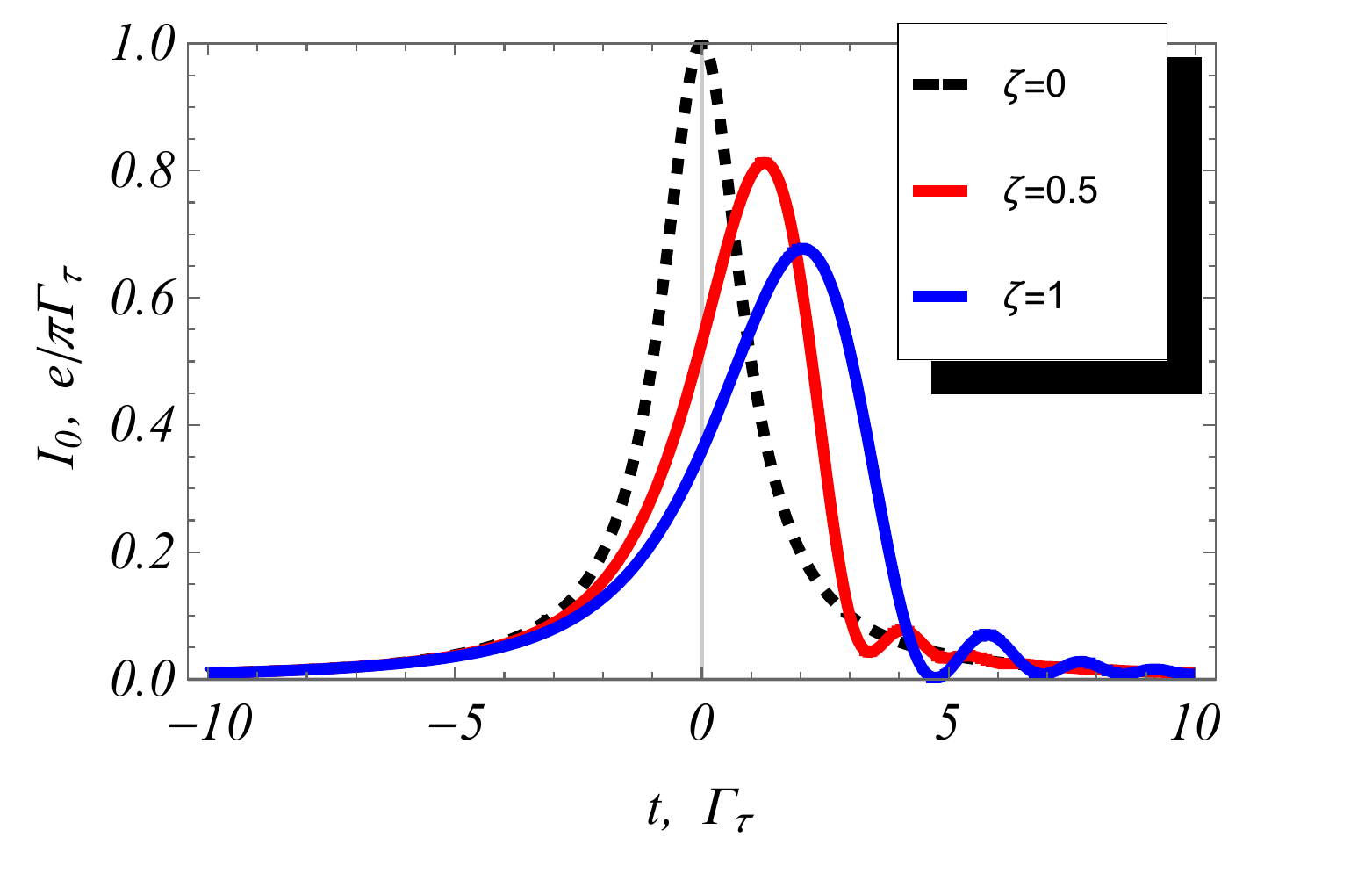} 
\includegraphics[width=75mm]{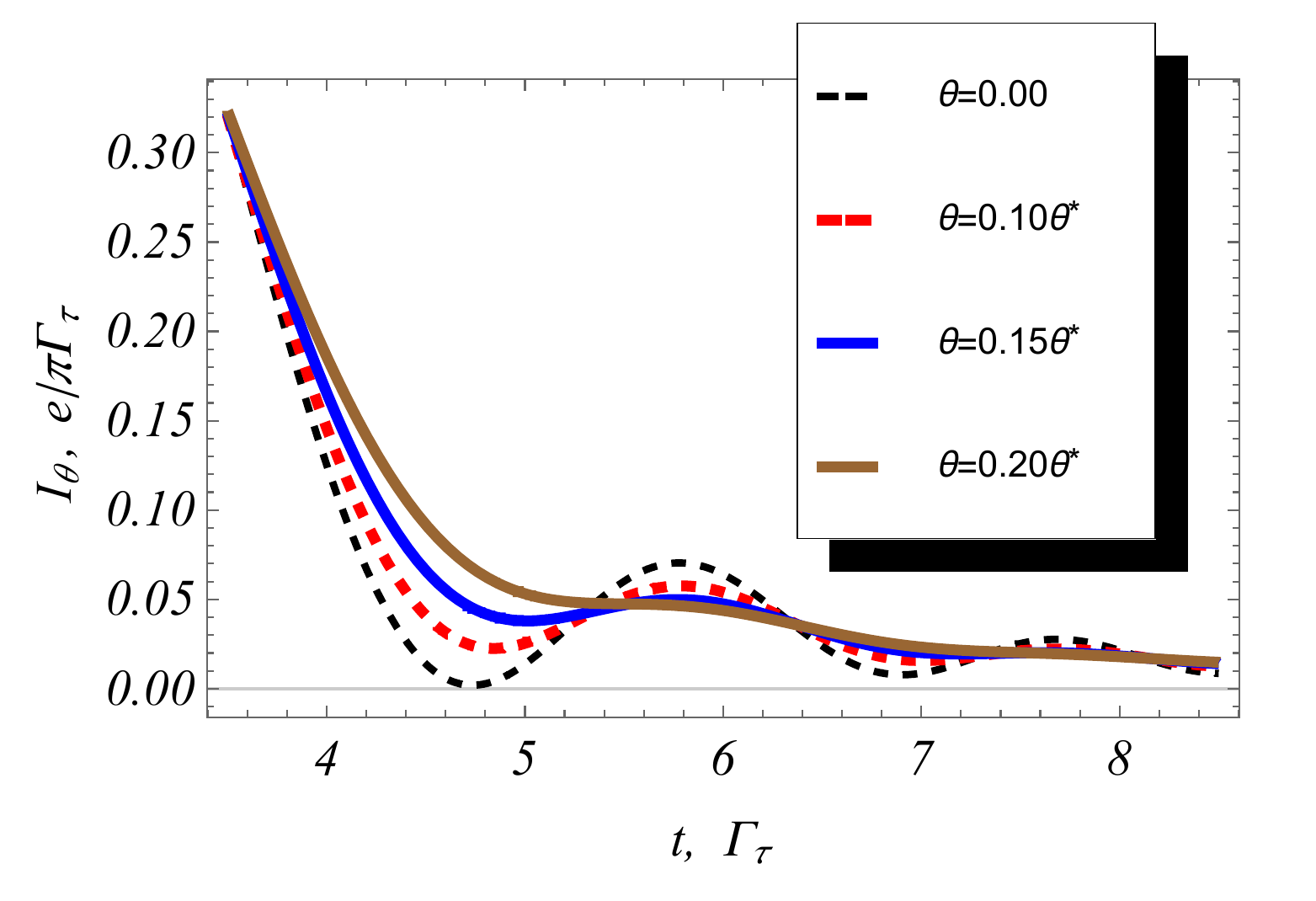}
\caption{(Color online) 
{\bf Left panel:} A time-dependent electrical current at zero ambient temperature $I_{0}(t)$, Eq.~(\ref{e03}), is shown for the adiabaticity parameter $ \zeta = 0$, $ \zeta= 0.5$, and $ \zeta=  1$  (in the order of decreasing maximum).
{\bf Right panel:} A time-dependent electrical current $I_{ \theta}(t)$, Eqs.~(\ref{e02}) and (\ref{e03}), is shown for different temperatures, $ \theta = 0$, $ \theta= 0.1 \theta^{*}$, $ \theta=  0.15 \theta^{*}$, and $ \theta=0.2 \theta^{*}$ in the order of decreasing oscillations. The effective temperature $ \theta^{*} = \hbar/ ( k_{B} \Gamma _{\tau})$.  The adiabaticity parameter $ \zeta = 1.0$. 
}
\label{It0}
\end{figure*}

At zero temperature an electron state is described by the wave function $ \Psi ^{(c)}(t)$, Eq.~(\ref{ex02}).\cite{Keeling:2008ft} 
Using Eq.~(\ref{ecf03}) for a zero-temperature correlation function we find a corresponding current,\cite{Moskalets:2016va}
\begin{eqnarray}
I_{0}(t) &=& 
e v _{ \mu} \left | \Psi ^{(c)} (t) \right | ^{2} = 
\frac{ e }{  \pi  \Gamma _{\tau} } 
\iint _{0}^{ \infty } \frac{ d \epsilon  d \epsilon^{\prime}  }{ 4 \epsilon _{0} ^{2}  }
e^{ - \frac{ \epsilon + \epsilon^{\prime}  }{2 \epsilon _{0}  } } 
\nonumber \\
 \label{e03} \\
&&\times
\cos  \left[ \frac{  \left( \epsilon^{\prime} - \epsilon \right) t  }{ \hbar  } +  \zeta \frac{ \epsilon^{2} - \left( \epsilon^{\prime}  \right)^{2} }{4 \epsilon _{0} ^{2} }  \right] 
.
\nonumber 
\end{eqnarray}
\ \\ \noindent
This current is shown in Fig.~\ref{It0}, the left panel for several values of the  adiabaticity parameter $ \zeta$. 

In the adiabatic emission regime,  $\zeta =0$, when the energy of a level changes very slow, $c \to  0$, an  electrical current $I_{0}(t)$ is a smooth function of time. 
With increasing rapidity (and, therefore, with increasing the non-adiabaticity parameter, since  $ \zeta \sim c$) some oscillations develop. 
These oscillations demonstrate redistribution of an electron density probability in time due to interference of amplitudes describing the emission process. 

To understand why oscillations vanish in the adiabatic emission regime and are present in the non-adiabatic emission regime we need to recall two time scales important to our problem.

The first one is {\it the dwell time} $ \tau _{D}$, which characterizes the duration of escape of a quantum state with given energy.\cite{{Wigner:1955us},{Smith:1960tm}} 
The initial state of an electron in the dot is a superposition of states with energies within the interval $2 \delta$. 
As we already mentioned, the time, after which the unoccupied states become available outside,  is different for different components of this superposition. 
Therefore, the different components of the initial superposition start to escape at different times and their escape leasts $ \tau _{D}$. 

The second important time is {\it the crossing time} $ 2\Gamma _{\tau} = 2 \delta / c$ -- a time during which the quantum level of width $2 \delta$ crosses the Fermi level. 
Let us emphasize the difference between the dwell time $ \tau _{D}$ and the crossing time $ \Gamma _{\tau}$:  
The crossing time characterizes an extent in time of the entire emitted state, which is a superposition of states with different energies. 
In contrast, the dwell time characterizes how fast a component of the initial  wave function with definite energy escapes the dot. 

The adiabatic emission regime is realized when $\Gamma _{ \tau} \gg \tau _{D}$. \cite{Splettstoesser:2008gc,Moskalets:2013dl}
In this case the energy of a level almost does not change on the scale of the dwell time.
Therefore, the components of the wave function, which have different energies before escape, so to say, leave the dot at different times and, therefore, do not overlap after the escape. 
No interference in time pattern arises. 
In the adiabatic emission regime the time profile of an electrical current $I _{0}(t)$, (see Fig.~\ref{It0}, the left panel, a black dashed line) reflects merely the Breit-Wigner density of states profile of a quantum  dot level (the rapidity $c$ plays a role of the transformation factor). 

With increasing rapidity $c$, the crossing time $ 2\Gamma _{\tau} = 2\delta / c$, decreases. 
When $\Gamma _{ \tau}$ becomes of the order of  the dwell time, $\Gamma _{ \tau} \sim \tau _{D}$, a non-adiabatic regime of emission  is established. 
The main feature of this regime is that the escape dynamics on the time-scale of the dwell time, $ \tau _{D}$, becomes essential. 

To clarify the escape dynamics in a non-adiabatic regime, let us take a  component of initial  wave function with some energy ${\cal E}$. 
It starts to escape at time $t _{0}$ (when ${\cal E}(t _{0}) = \mu$) and (almost) finishes to escape at $t _{0} + \tau _{D}$. 
However, after time $ \tau _{D}$, the energy of a quantum level increases by a noticeable  amount of $c \tau _{D} \sim c \Gamma _{\tau} \sim \delta$. 
As a result, many components of the  initial wave function -- namely those with energy  $ {\cal E} _{0} - c \tau _{D} < {\cal E} < {\cal E} _{0}$ -- are already started to escape. 
So, since the escape process takes a finite time, $ \tau _{D}$,  there are many components of the wave function that are in the process of escaping and that are overlapping at the exit of an electron source.
This overlap causes interference. 

Let us repeat, the partial states interfering at a given time at the exit of a source started to escape at different times. 
While the spatial places, where the escape starts and ends, are the same for  all partial states. 
Therefore, the interfering states differ by the paths in time they took to escape the source. 
This is why I use a term {\it interference in time}.  

Interference in time results in time-oscillations of physical quantities. 
For instance, a time-dependent current $I(t)$, shows such oscillations, see Fig.~\ref{It0}.   
These oscillations are noticeable at later times -- positive times on Fig.~\ref{It0} --, when the intensity of what remains to escape (from the states, which started to escape earlier) is of the order of the intensity of states that are starting to escape.

The scenario outlined above  remains valid for finite temperatures as well.
Therefore, the oscillations caused by interference in time can be also observed at non zero temperatures, 
see Fig.~\ref{It0}, the right panel. 
However the amplitude of oscillations decreases with temperature. 
The explanation is the following. 
At finite temperatures, as we already mentioned, the pure state becomes a mixed state, whose components are more spread in energy. 
This enhanced spread in energy is accompanied by an enhanced spread in time that diminishes oscillations. 

Naturally the energy of an emitted electron sets a scale for temperature.  
In the case of a quantum level raising at a constant rapidity, see Eq.~(\ref{ex02}), we define the characteristic temperature as $ k_{B}\theta^{*} = 2 \epsilon_{0} \Rightarrow \theta^{*} = \hbar/ (k_{B}  \Gamma _{\tau}) $. 
From Fig.~\ref{It0}, the right panel we find that the oscillations are noticeable if $\theta < 0.1 \theta^{*}$. 
Therefore, to observe oscillations of a time-dependent electrical current due to interference in time for electronic temperature $ \theta = 30$\,mK the width of a wave-packet should be not larger than $2  \Gamma _{\tau} < 0.2 \hbar / (k_{B} \theta) \approx 5\times 10^{-11}\,\mathrm{s} = 50\,\mathrm{ps}$.

\subsection{Time-dependent heat current}

Interference in time has even more striking effect on a heat current.
While a single-particle electrical current is positive definite at any time, $I(t) = e v _{ \mu} \left |  \Psi(t) \right | ^{2}$ [we used Eqs.~(\ref{e01}) and (\ref{ecf03})], the  heat current has no such a constrain. 
In terms of the correlation function the heat current reads,\cite{Moskalets:2016va} 

\begin{eqnarray}
I^{Q}(t) =
v_{ \mu} \left[ \left\{ \frac{ -i \hbar }{ 2} \left( \frac{ \partial }{ \partial t } - \frac{ \partial }{ \partial t^{\prime} } \right) -  \mu \right\} G^{(1)}_{}(t;t^{\prime}) \right] _{t = t^{\prime}} . 
\label{h01}
\end{eqnarray}
\ \\ \noindent
For a single-particle state it becomes, $I^{Q}(t) = v_{ \mu} \hbar \, {\rm Im} \left [ \frac{ \partial \psi^{*}(t) }{ \partial t } \psi(t) \right]$, which indeed does not look like positive definite. 
This fact illustrates that a heat current is fundamentally different from a charge current not only for interacting systems, see, e.g. Ref.~\onlinecite{Banerjee:2016tz}, but even for the system on non-interacting fermions, see, e.g. Ref.~\onlinecite{{Ludovico:2014de},{Rossello:2014vb},Ronetti:2016un}. 

The heat current caused by injection of a  single electron with wave function $ \Psi ^{(c)}$, Eq.~(\ref{ex02}), was analyzed in Ref.~\onlinecite{Moskalets:2016va}. 
It was rather a surprise, that at intermediate values of the adiabaticity parameter $ \zeta$ -- when interference in time manifests itself the most --  the heat current can become negative for short times. 
A negative heat current appeared also in calculations of Refs.~\onlinecite{Ludovico:2014gq,{Ludovico:2016hh}}. 
In this context negative means that a heat current is directed not from the electronic source into a zero-temperature reservoir but back.

This fact puts in focus the issue of interpretation of a heat current in quantum systems and, in particular, of a heat current associated to a single-particle excitation.

\begin{figure*}[t]
\includegraphics[width=80mm]{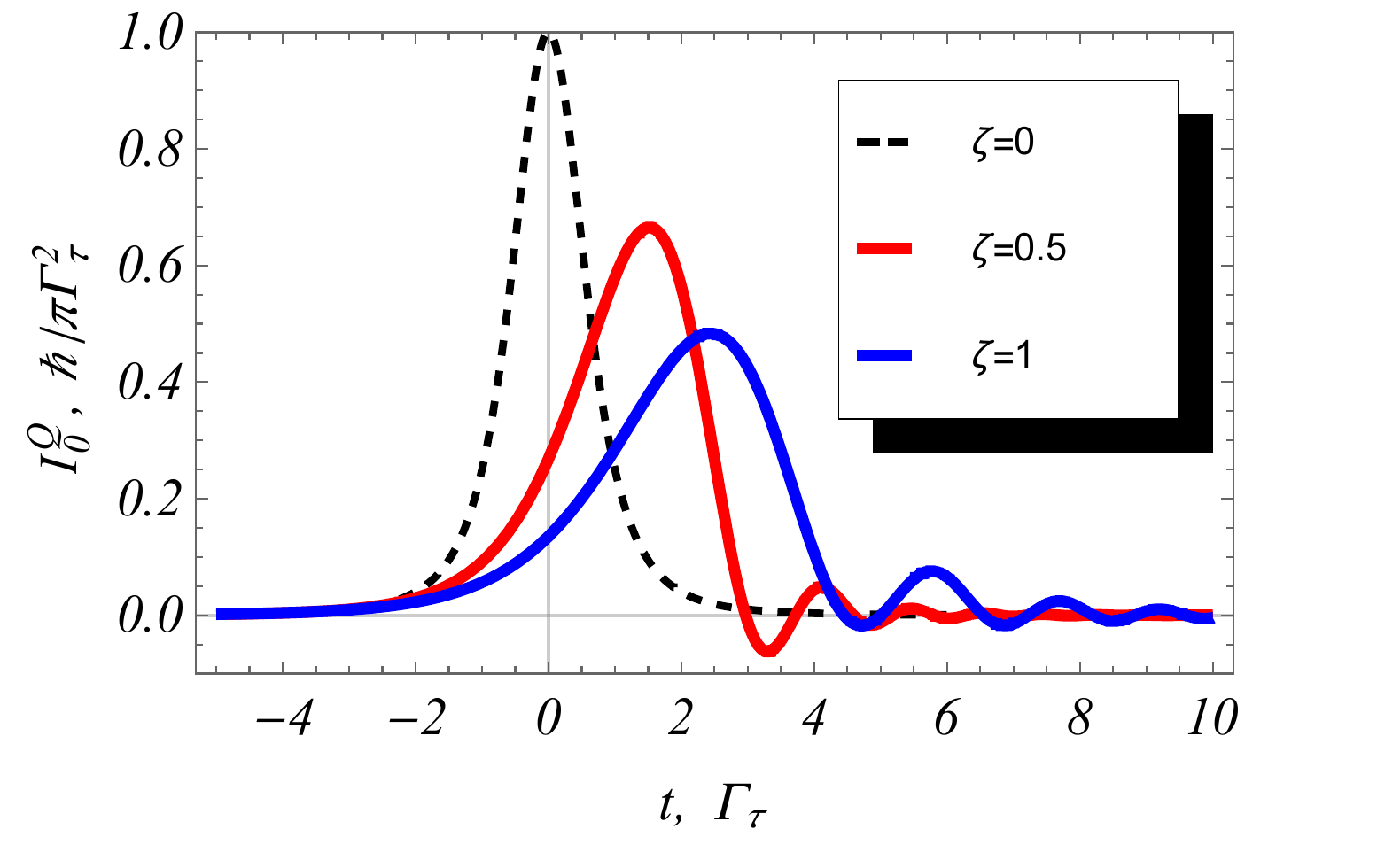}
\includegraphics[width=80mm]{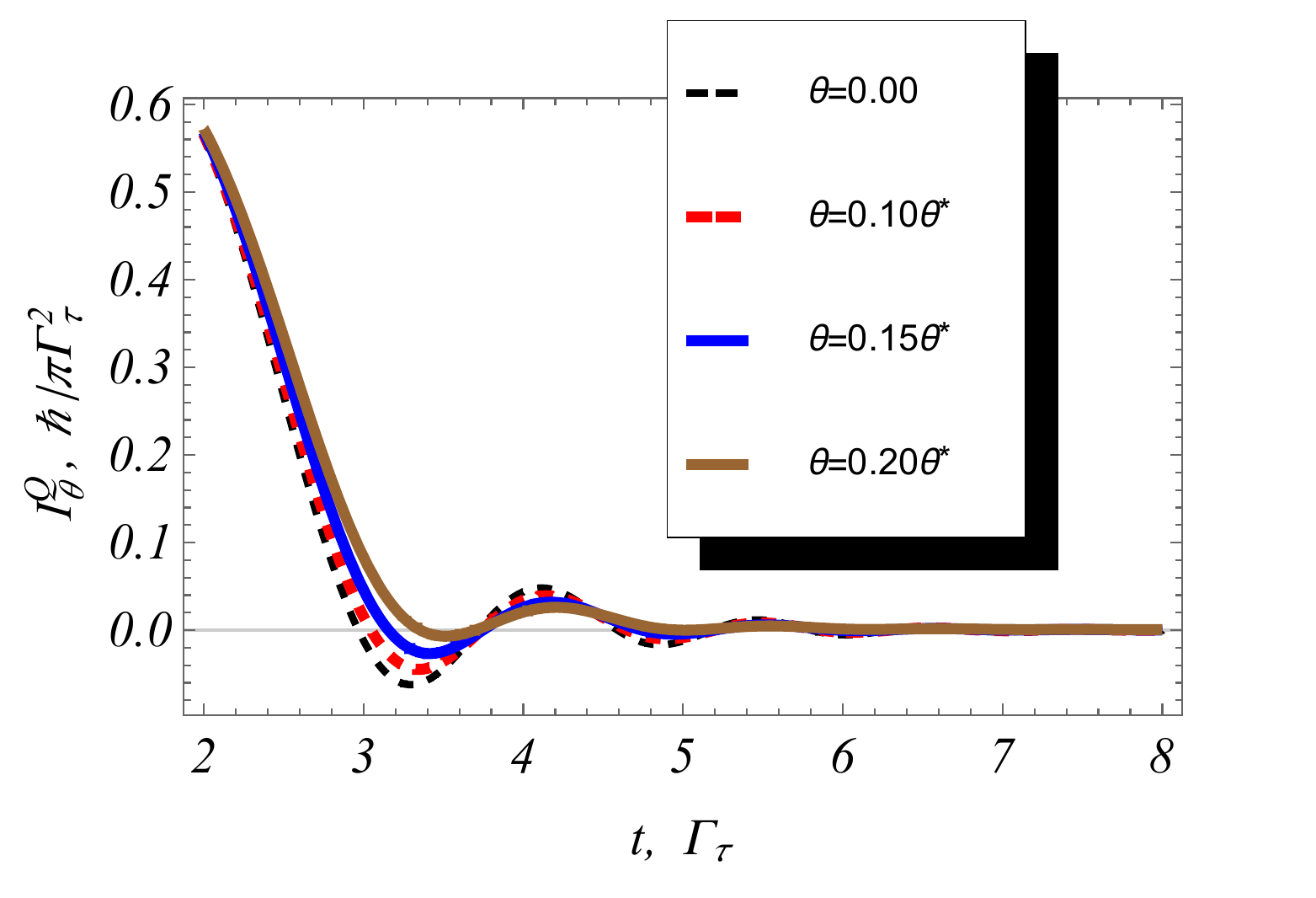}
\caption{
(Color online) {\bf Left panel:} A time-dependent heat current $I^{Q}_{0}(t)$ at zero ambient temperature, Eq.~(\ref{h03}), is shown for different values of the adiabaticity parameter, $ \zeta = 0$, $ \zeta= 0.5$, and $ \zeta=  1$ shown in the order of decreasing maximum. 
{\bf Right panel:} A time-dependent heat current $I^{Q}_{ \theta}(t)$, Eq.~(\ref{h02}) with $I _{0}$ from Eq.~(\ref{e03}) and $I ^{Q} _{0}$ from Eq.~(\ref{h03}), is shown for different temperatures, $ \theta = 0$, $ \theta= 0.1 \theta^{*}$, $ \theta=  0.15 \theta^{*}$, and $ \theta=0.2 \theta^{*}$ in the order of decreasing oscillations. The effective temperature $ \theta^{*} = \hbar/ ( k_{B} \Gamma _{\tau})$.  The adiabaticity parameter $ \zeta = 0.5$.}
\label{Qt0}
\end{figure*}

The na\"{i}ve classical-like interpretation of a heat current direction as the  direction where heat runs to is completely unphysical. 
In fact, the heat current characterizes how heat content of a quantum system changes in time. 
In the case of a freely moving singe particle under consideration here, the heat current gives the rate of change of (a strictly positive) heat carried by such a particle. 
A negative heat current means that heat decreases with time. 
In other words, for those times, when a heat current is negative, the measurement performed at later times would reveal lesser (but still positive) heat carried by a particle.   

This sound counterintuitive. 
Indeed, due to the energy conservation the heat is nothing but the work done by a dynamic force driving the source. 
It is natural to expect that this force performs more work when it acts longer on an emitted particle. 
On the other hand, when we perform a measurement on a particle, we break the connection between a particle and a driving force. 
As a result, an earlier measurement is expected to show a lesser heat associated to a particle. 

A negative heat current, which is counterintuitive according to classical expectations, is a result of quantum-mechanical interference. 
Therefore, a negative heat current witnesses a quantum state that has no underlying classical model. 

Here I explore whether a negative heat current persists at finite temperatures. 
Let us use the general equation (\ref{ecf08}) and express a finite-temperature heat current $I^{Q}_{ \theta}(t)$ in terms of a zero-temperature heat current $I^{Q}_{0}(t)$ and a zero-temperature charge current $I_{0}(t)$, 

\begin{eqnarray}
I^{Q}_{ \theta}(t) &=& 
\int\limits_{- \infty}^{ \infty}    d \epsilon ^{\prime}
\left( - \frac{  \partial f( \epsilon ^{\prime} ) }{  \partial \epsilon ^{\prime} } \right) 
I^{Q}_{0}\left ( t -  \frac{ \epsilon ^{\prime} }{ c  } \right) 
\nonumber \\
\label{h02} \\
&& +
\frac{ 1 }{ e  }
\int\limits_{- \infty}^{ \infty}    d \epsilon ^{\prime} 
\left( - \frac{  \partial f( \epsilon ^{\prime} ) }{  \partial \epsilon ^{\prime}  } \right) 
\epsilon ^{\prime}
I_{0}\left ( t - \frac{  \epsilon ^{\prime} }{ c  } \right) .
\nonumber 
\end{eqnarray}
\ \\ \noindent
Note that the second term in Eq.~(\ref{h02}) comes from the phase factor in Eq.~(\ref{ecf08}), which reflects renormalization of the effective Fermi energy for different components of the mixed state at finite temperature, see Eq.~(\ref{ecf09}). 

Likewise as an electrical current, the heat carried by a leviton is not affected by finite temperatures, $I _{ \theta} ^{Q(L)}(t) = I _{0} ^{Q(L)}(t)$,  see a remark after Eq.~(\ref{e02}).

\subsubsection{Single quantum level raising at a constant rapidity}

The wave function $ \Psi ^{(c)}(t)$, Eq.~(\ref{ex02}), substituted into Eq.~(\ref{ecf03}) and then into Eq.~(\ref{h01}) results in a zero-temperature heat current,

\begin{eqnarray}
I^{Q}_{0}(t) &=& 
\frac{ \hbar }{ \pi  \Gamma _{\tau} ^{2} } 
\iint _{0}^{ \infty } \frac{ d \epsilon  d \epsilon^{\prime}   }{ 4  \epsilon _{0} ^{2} }
\frac{ \epsilon + \epsilon^{\prime} }{ 2  \epsilon _{0}  }  
e^{ - \frac{ \epsilon + \epsilon^{\prime}  }{2 \epsilon _{0} }  } 
\nonumber \\
\label{h03} \\
&&\times
\cos  \left[  \frac{ \left( \epsilon^{\prime} - \epsilon \right) t }{ \hbar  }  +  \zeta \frac{ \epsilon^{2} - \left( \epsilon^{\prime}  \right)^{2} }{4  \epsilon _{0} ^{2} }  \right]  
.
\nonumber 
\end{eqnarray}
\ \\ \noindent
This current is shown in Fig.~\ref{Qt0}, the left panel for different values of the adiabaticity parameters $ \zeta$. 
In the adiabatic emission regime,  $\zeta =0$ the heat current $I_{0}^{Q}(t)$ is a smooth and positive function of time. 
As we discussed above, no interference in time occurs in this regime. 

With increasing $ \zeta$ the heat current becomes oscillating in time. 
Since these oscillations are due to interference (interference in time), their existence is a manifestation of the quantum nature of a carrier. 
A distinctive feature of heat current oscillations is that they fall below zero.
This fact has no classical explanation. 
With increasing temperature, when a pure state becomes a mixed state, the amplitude of oscillations and the time interval, where heat is negative,   are diminished, see Fig.~\ref{Qt0}, the right panel.

\section{Conclusion}
\label{concl}

I found the general condition, which guarantees that an electron source, which  injects single particles onto the surface of the Fermi sea at zero temperature, works as a single-particle source at finite temperatures as well. 
According to this condition the scattering amplitude of a source, $S _{in}(t, \epsilon)$, has to possess the time-energy translation symmetry, $S _{in}(t , \epsilon) = S _{in}(t  - \delta \epsilon /c, \epsilon - \delta \epsilon)$, Eq.~(\ref{ecf07}), where $c$ is a constant.  
In this case the temperature effect boils down to the fact that the injected  single-particle state becomes a mixed state whose components have the same in shape but shifted in time envelope functions, see Eq.~(\ref{ecf09}). 
The probability density is given by the energy derivative of the Fermi distribution function.  

A particular case, when $S _{in}$ is energy independent, falls into this category as well.   
The recently realized source of levitons, see Ref.~\onlinecite{Dubois:2013ul}, is  characterized by an energy-independent scattering amplitude. 
Therefore, at finite temperatures it emits a single-particle mixed state.   

I analyzed a few theoretical models of a single-electron source  whose scattering amplitude does possess the time-energy translation symmetry.   
In particular, under quite general conditions a single-electron source based on a  quantum capacitor of Ref.~\onlinecite{Feve:2007jx} is described by such a model. 
To be specific, it is described by the model of a single quantum level raising at a constant rapidity, which was put forward in Ref.~\onlinecite{Keeling:2008ft}. 
An analytical expression for the wave function  provided by this model allows to analyze in detail the properties of states produced at adiabatic as well as at non-adiabatic emission regimes.  

The non-adiabatic emission regime of a quantum capacitor driven by a fast harmonic potential is particularly interesting. 
In this regime the interference of partial amplitudes corresponding to processes that last different time (the interference in time) manifests itself in oscillations of time-dependent electrical and heat currents associated with a single-particle excitation. 
Moreover, a time-dependent heat current can temporarily even change a sign, that is forbidden from the classical point of view. 
Therefore, a time-resolved  heat current (when it is negative) can witness a quantum state that has no underlying classical model. 

The last finding reveals two interesting things. 
First, the transport measurements can be used instead of the Wigner function measurement \cite{Ferraro:2013bt} to provide evidence of  quantum states with no classical interpretation. 
Second, the single-electron source can be used to generate such quantum states.


\appendix

\section{A harmonically driven quantum capacitor: The Floquet scattering matrix formalism}
\label{appA}

In this appendix we derive the scattering amplitude $S ^{(c)} _{in}$, Eq.~(\ref{ex01}), found in Ref.~\onlinecite{Keeling:2008ft} as the limiting case of the scattering amplitude of a quantum capacitor $S ^{(cap)} _{in}$, Eq.~(\ref{ex04}), found in Ref.~\onlinecite{Moskalets:2008fz}. 

First, we use the wide band approximation and linearize $k(E)$ in Eq.~(\ref{ex04}) around the Fermi energy $ \mu$. 
In this case the kinematic phase $ \varphi(E) = k(E) L$ becomes, $ \varphi(E) = \varphi _{ \mu} + \tau \hbar ^{-1} \epsilon$, where $ \varphi _{ \mu}$ is a  phase calculated at the Fermi energy,  $ \tau = L / v( \mu)$ is the time of a single revolution calculated for an electron with Fermi energy, and $ \epsilon = E - \mu$ is an electron energy counted from the Fermi energy $ \mu$. 

Second, we go to the limit of a large level spacing, $ \Delta = h/ \tau$  (to unsure that only one level contributes to emission), while keeping the dwell time finite, $ \tau_{D} = \tau/T$: (i) $ \Delta \to \infty \Rightarrow  \tau\to 0$, (ii)  $ 0 < \tau/T  \Rightarrow T \to 0$. 
Remind that $T$ is the transmission probability of a QPC connecting the dot and the waveguide. 

Third, we suppose that the amplitude of a driving potential $U(t) = U _{0} \cos \left( \Omega t \right)$ is of the order of the level spacing, $eU _{0} \sim \Delta$. 
Therefore, since $ \Delta \gg \delta$, the time interval $  2\Gamma _{\tau}$, during which a quantum level of width $2 \delta = T \Delta/ (2 \pi) = \hbar/ \tau_{D}$ crosses the Fermi level, is small compared to the half-period ${\cal T}/2 = \pi/ \Omega$, during which the energy of a level changes by $2U _{0} \sim \Delta$. 
If so, we can linearize $U(t)$ close to time $t _{0}$, when a quantum level crosses the Fermi level: 
$eU(t) \approx eU(t _{0}) + c (t - t _{0})$, where the rapidity $c = edU/dt$ is calculated at $t = t _{0}$. 
Without loss of generality we put $t _{0} = 0$. 

The approximation of a constant rapidity allows us to simplify a time-dependent phase, $ \Phi _{q}$, Eq.~(\ref{ex05}), as follows,

\begin{eqnarray}
\Phi_{q}(t)=q \Phi _{0} + \frac{c}{\hbar}\left( t q \tau - \frac{q^{2} \tau^{2} }{2 }  \right) ,
\label{lt01}
\end{eqnarray}
\ \\ \noindent
where $ \Phi _{0} = e U _{0} \tau / \hbar$. 
In the equation above the first term, $q \Phi _{0}$, is a phase picked up by an electron during $q$ revolutions at a constant potential $U _{0}$. 
The second term accounts for the change of potential in time.  

Using these approximations we rewrite the scattering amplitude in Eq.~(\ref{ex04}) as follows [$ S _{in} ^{(cap)}\left(  t, \mu + \epsilon \right) \equiv S _{in} ^{(cap)}\left(  t, \epsilon \right)$],

\begin{eqnarray}
S_{in} ^{(cap)}(t, \epsilon) = \sqrt{R} e ^{i \theta _{r}} \left\{ 1 - \frac{T}{R}    \sum_{q=1}^{\infty} \left(  \sqrt{R} \right)^{q} e^{i q \Theta} \right\}, 
\label{lt02}
\end{eqnarray}
where 

\begin{eqnarray}
\Theta = \Theta _{0}  +  \frac{  \tau }{ \hbar } \left(  \epsilon - c t  \right)  + \frac{ c q \tau ^{2} }{ 2 \hbar  },
\label{lt03}
\end{eqnarray}
\ \\ \noindent
is the total phase accumulated during a single revolution and $ \Theta _{0} =  \theta _{r} + \varphi_{ \mu} - \Phi _{0}$. 
The constant contribution $ \Theta _{0}$ can be eliminated if we redefine the origin of time. 
In what follows we put $ \Theta _{0} = 0$, which means that a raising quantum  level crosses the Fermi level at $t = 0$. 

In equation (\ref{lt02}) we introduced the modulus $R = 1 - T$ and the phase $ \theta _{r}$ of a reflection amplitude, $r = \sqrt{R} e ^{i \theta _{r}}$. 
In addition we use $\tilde t/r = - \tilde t ^{*}/r ^{*}$, which follows from the unitarity of the scattering matrix of a QPC connecting a capacitor and a waveguide.

Next, we replace the sum over the number of revolutions $q$ in Eq.~(\ref{lt02}) by the corresponding integral. 
We can do this, since the time of a single revolution $ \tau$ is small compared to other relevant time scales, such as the dwell time $ \tau _{D}$, the crossing time $ 2\Gamma _{\tau} = 2\delta / c$, etc.  
We introduce $ \xi = q T \equiv q\tau / \tau_{D}$ and replace 

\begin{eqnarray}
T \sum\limits_{q=1}^{\infty} \to \int _{0}^{ \infty } d \xi .
\label{lt04}
\end{eqnarray}
\ \\ \noindent

Finally, since $T \to 0$ we can transform $ \left(  \sqrt{R}  \right) ^{q} = e^{ \frac{q}{2} \log(1-T)} \approx e^{- q\frac{ T }{ 2 }} = e ^{- \frac{ \xi }{ 2 }}$. 
In the two additional factors, in front of the sum and in front of the curly brackets in Eq.~(\ref{lt02}), we put $R \approx 1$ and obtain

\begin{eqnarray}
S_{in} ^{(cap)}(t, \epsilon) =  
e ^{i \theta _{r}} 
\left\{  
1-  \int\limits _{0}^{ \infty } d \xi 
e^{ - \frac{ \xi }{2 } } 
e^{ \frac{ i }{ \hbar  } \xi  \tau _{D}  \left(  \epsilon - c t \right) } 
e^{  i \xi^{2}  \frac{\zeta }{ 4}   }
\right\}
, 
\label{lt05}
\end{eqnarray}
\ \\ \noindent
where we introduced the parameter $ \zeta = 2 c \tau _{D} ^{2} / \hbar $. 

Apparently the equation above in nothing but the equation (\ref{ex01}) up to the irrelevant phase factor $e ^{i \theta _{r}}$.

\subsection{Unitarity}

The scattering amplitude $S _{in} ^{(cap)}(t,E)$, Eq.~(\ref{ex04}), is unitary. 
The proof is rather simple and it can be found, for instance, in Ref.~\onlinecite{Moskalets:2011wz}. 
Let us check the unitarity of the transformed amplitude to be sure that we missed nothing essential on the way from equation (\ref{ex04}) to equation (\ref{lt05}).

The unitarity condition for the scattering amplitude of a periodically driven  scatterer in the mixed energy-time representation reads, \cite{{Moskalets:2008ii}}

\begin{eqnarray}
\int _{ - \frac{ {\cal T} }{ 2 }}^{\frac{ {\cal T} }{ 2 } } \frac{dt}{{\cal T}} \left | S _{in} ^{(cap)} \left(  t,E \right) \right | ^{2} = 1, 
\label{lt06}
\end{eqnarray}
\noindent \\
where ${\cal T}$ is the period of drive. 

When we went over from the sum over $q$ in Eq.~(\ref{ex04}) to the integral over $ \xi$ in Eq.~(\ref{lt05}), we supposed that the amplitude of a driving potential is large, $eU _{0} \sim \Delta \to \infty$. 
Together with the assumption of a finite rapidity, $c < \infty$, this means formally that the period of a drive is large, ${\cal T} \to \infty$. 
Therefore, we rewrite the equation above as follows,

\begin{eqnarray}
\lim _{ {\cal T} \to \infty} \int _{ - \frac{ {\cal T} }{ 2 }}^{\frac{ {\cal T} }{ 2 } } \frac{dt}{{\cal T}} 
\left | S _{in}  ^{(cap)} \left(  t,E \right) \right | ^{2} - 1 =
\quad
\label{lt07} \\
=
\lim _{ {\cal T} \to \infty} \int _{ - \frac{ {\cal T} }{ 2 }}^{\frac{ {\cal T} }{ 2 } } \frac{dt}{{\cal T}} 
\left\{  
1-  \int\limits _{0}^{ \infty } d \xi 
e^{ - \frac{ \xi }{2 } } 
e^{ \frac{ i }{ \hbar  } \xi  \tau _{D}  \left(  \epsilon - c t \right) } 
e^{  i \xi^{2}  \frac{\zeta }{ 4}   }
\right\}
\nonumber \\
\times
\left\{  
1-  \int\limits _{0}^{ \infty } d \xi 
e^{ - \frac{ \xi }{2 } } 
e^{ - \frac{ i }{ \hbar  } \xi  \tau _{D}  \left(  \epsilon - c t \right) } 
e^{  -i \xi^{2}  \frac{\zeta }{ 4}   } 
\right\} - 1 = A + B, 
\nonumber 
\end{eqnarray}
\noindent \\
where 

\begin{eqnarray}
A = - 2 {\rm Re} \left\{  
\lim_{ {\cal T}\to \infty } \int _{- \frac{ {\cal T} }{2 }}^{ \frac{ {\cal T} }{2 }} \frac{dt }{ {\cal T}  } 
\int\limits _{0}^{ \infty } d \xi 
e^{ - \frac{ \xi }{2 } } 
e^{  \frac{ i }{ \hbar  } \xi  \tau _{D}  \left(  \epsilon - c t \right) } 
e^{  i \xi^{2}  \frac{\zeta }{ 4}   } 
\right\} ,
\nonumber \\
\label{lt08}
\end{eqnarray}

\begin{eqnarray}
B = \lim_{ {\cal T}\to \infty } \int _{- \frac{ {\cal T} }{2 }}^{ \frac{ {\cal T} }{2 }} \frac{dt }{ {\cal T}  } 
\int\limits _{0}^{ \infty } d \xi 
e^{ - \frac{ \xi }{2 } } 
e^{  \frac{ i }{ \hbar  } \xi  \tau _{D}  \left(  \epsilon - c t \right) } 
e^{  i \xi^{2}  \frac{\zeta }{ 4}   } 
\nonumber \\
\times 
\int\limits _{0}^{ \infty } d \xi ^{\prime} 
e^{ - \frac{ \xi ^{\prime} }{2 } } 
e^{ - \frac{ i }{ \hbar  } \xi ^{\prime} \tau _{D}  \left(  \epsilon - c t \right) } 
e^{  -i \left(  \xi ^{\prime} \right)^{2}  \frac{\zeta }{ 4}   } . 
\label{lt09} 
\end{eqnarray}
\noindent \\
We need to show that $A + B = 0$. 

First of all we integrate out $t$ in $A$,

\begin{eqnarray}
\lim_{ {\cal T}\to \infty } 
\int _{- \frac{ {\cal T} }{2 }}^{ \frac{ {\cal T} }{2 }} \frac{dt }{ {\cal T}  } 
e^{-i t \xi \frac{ c \tau _{D} }{ \hbar} }
=  \lim_{ {\cal T}\to \infty } \frac{2 \pi }{ {\cal T} } \frac{ \hbar }{ c \tau _{D}  } \delta\left( \xi \right),
\label{lt10} 
\end{eqnarray}
\noindent \\
and in $B$,

\begin{eqnarray}
\lim_{ {\cal T}\to \infty } 
\int _{- \frac{ {\cal T} }{2 }}^{ \frac{ {\cal T} }{2 }} \frac{dt }{ {\cal T}  } 
e^{i t \left( \xi ^{\prime} - \xi \right) \frac{ c \tau _{D} }{ \hbar} }
=  
\lim_{ {\cal T}\to \infty } 
\frac{2 \pi }{ {\cal T} } 
\frac{ \hbar }{ c \tau _{D}  } 
\delta\left( \xi ^{\prime} - \xi \right). 
\nonumber \\
\label{lt11}
\end{eqnarray}

Then we evaluate $A$, Eq.~(\ref{lt08}). 
We substitute Eq.~(\ref{lt10}) in Eq.~(\ref{lt08}), evaluate the integrand at $ \xi=0$, use $ \int _{0}^{ \infty } d \xi \delta ( \xi) = \frac{1}{2}$, and get,

\begin{eqnarray}
A = - 2 \lim_{ {\cal T}\to \infty }  \frac{2 \pi }{ {\cal T} } 
\frac{ \hbar }{ c \tau _{D}  } \frac{1}{2} .
\label{lt12}
\end{eqnarray}

To evaluate $B$ we substitute Eq.~(\ref{lt11}) into Eq.~(\ref{lt09}). 
Then we use $ \delta\left(  \xi ^{\prime} - \xi \right)$ and integrate out $ \xi ^{\prime}$. 
Since $ \xi ^{\prime} = \xi$, the oscillating factors cancel each other. 
After evaluation of the remaining integral, $ \int _{0}^{ \infty } d \xi e ^{- \xi} = 1$, we find

\begin{eqnarray}
B = 
\lim_{ {\cal T}\to \infty } 
\frac{2 \pi }{ {\cal T} } 
\frac{ \hbar }{ c \tau _{D}  } .
\label{lt13}
\end{eqnarray}
\noindent \\ 
Comparing Eqs.~(\ref{lt12}) and (\ref{lt13}) we see that indeed $A + B = 0$, as expected. 
So, the scattering amplitude $S _{in}  ^{(cap)}$, Eq.~(\ref{lt05}) does satisfy the unitarity condition Eq.~(\ref{lt06}).

\subsection{Single-electron wave function}

For completeness, let us derive the wave function $ \Psi ^{(c)}$, Eq.~(\ref{ex02}), directly from the excess correlation function $G ^{(1)}$, Eq.~(\ref{ecf01}), using the scattering amplitude $S _{in} ^{(cap)}$, Eq.~(\ref{lt05}). 
Note that originally the wave function $ \Psi ^{(c)}$ was calculated in Ref.~\onlinecite{Keeling:2008ft} using another approach. 

We substitute Eq.~(\ref{lt05}) into Eq.~(\ref{ecf01}) and replace  $ \int _{- \infty}^{ \infty } d \epsilon f(E) \to \int _{ - \infty}^{ 0 } d \epsilon$ at zero temperature.  
Then, for the sake of convenience, we change $ \epsilon \to - \epsilon$ and go over to dimensionless variables. 
We normalize $t$ by $2  \Gamma _{\tau} = \hbar/ (c \tau _{D})$ and $ \mu$  and $ \epsilon = E - \mu$ by $ \epsilon_{0} = \hbar/( 2  \Gamma _{\tau})$. 
In addition we introduce the adiabaticity parameter $ \zeta =  \tau _{D}/ \Gamma _{\tau}$ and finally obtain (the subscript $ _{cap}$ indicates that we use the scattering amplitude of a quantum capacitor),

\begin{eqnarray}
G^{(1)} _{cap} (t_{1};t_{2}) = 
\frac{ e^{  i  \mu   \left( t_{1} - t_{2} \right)  }   }{  4 \pi  \Gamma _{\tau} v_{ \mu}  }
\sum\limits_{j=1}^{3} \gamma_{j} (t_{1};t_{2}) 
,
\label{lt16}
\end{eqnarray}
\ \\ \noindent
where 

\begin{subequations}
\label{lt17}

\begin{eqnarray}
 \gamma^{}_{1} (t_{1};t_{2}) &=& 
-
\int\limits _{0}^{ \infty } d \xi 
e^{ - \frac{\xi }{2 } } 
e^{ i \xi  t_{1}    } 
e^{ -i \xi^{2}  \frac{\zeta  }{ 4}  } 
\int\limits_{0}^{ \infty}    d \epsilon  
e^{ - i  \epsilon  t_{1}   } 
e^{ i  \epsilon t_{2}   } 
e^{ i \epsilon \xi \frac{\zeta }{ 2 }  } ,
\nonumber \\
\label{lt17a} 
\end{eqnarray}

\begin{eqnarray}
 \gamma^{}_{2} (t_{1};t_{2}) &=& 
-
\int\limits _{0}^{ \infty } d \xi 
e^{ - \frac{\xi }{2 } }
e^{ -i \xi  t_{2}   } 
e^{ i \xi^{2}  \frac{\zeta  }{ 4}  }  
\int\limits_{0}^{ \infty}    d \epsilon  
e^{ - i \epsilon  t_{1}  } 
e^{ i \epsilon t_{2}    } 
e^{ - i \epsilon \xi \frac{\zeta }{ 2 } } ,
\nonumber \\
\label{lt17b} 
\end{eqnarray}

\begin{eqnarray}
 \gamma^{}_{3} (t_{1};t_{2}) &=& 
\int\limits_{0}^{ \infty}    d \epsilon  
e^{ - i   \epsilon t_{1}   } 
e^{ i  \epsilon t_{2}  } 
\int\limits _{0}^{ \infty } d \xi^{\prime} 
e^{ - \frac{\xi^{\prime} }{2 } } 
e^{ i \xi^{\prime}  t_{1}   } 
e^{ -i \left( \xi^{\prime} \right)^{2}  \frac{\zeta  }{ 4}  } 
e^{ i \epsilon \xi^{\prime} \frac{\zeta }{ 2 }  } 
\nonumber \\
&& \times
\int\limits _{0}^{ \infty } d \xi 
e^{ - \frac{\xi }{2 } } 
e^{ -i \xi  t_{2}   } 
e^{ i \xi^{2}  \frac{\zeta  }{ 4}  } 
e^{ - i \epsilon \xi \frac{\zeta }{ 2 }  } 
.
\label{lt17c} 
\end{eqnarray}

\end{subequations}

To find a wave function we need to factorize Eq.~(\ref{lt16}) and then to use Eq.~(\ref{ecf03}). 
To this end we, first, need to get rid of the third integral in the equation above.

\subsubsection{Evaluation of $ \gamma _{3}$}

We make the following change of variables,  

\begin{eqnarray}
\epsilon &=&  
z + \frac{ \left | y \right | + y }{ 2 } + \frac{ \left |  x \right | - x }{ 2 }  
,
\nonumber \\
\xi^{\prime}  &=&  
z +  \frac{ \left |  x \right | +  x }{ 2 }
, \quad
\label{lt18} \\
\xi  &=& 
z +  \frac{ \left | y \right | - y }{ 2 } + \frac{ \left |  x \right | -  x }{ 2 }  ,
\nonumber \\
\nonumber 
\end{eqnarray}
\noindent \\
that also implies a change of the area of integration,

\begin{eqnarray}
\int\limits_{0}^{ \infty} d \epsilon 
\int\limits_{0}^{ \infty} d \xi^{\prime} 
\int\limits_{0}^{ \infty} d \xi 
&\to& 
\int\limits_{ 0 }^{ \infty} d  z 
\int\limits_{- \infty}^{ \infty} d  x 
\int\limits_{- \infty}^{ \infty} dy 
.
\label{lt19}
\end{eqnarray}
\noindent \\
To simplify notations we introduce,

\begin{eqnarray}
X_{+} &=& \frac{ \left | X \right | + X }{ 2 } = 
\left\{
\begin{array}{cc}
X,  & X>0, \\
\ \\
0,  & X<0, 
\end{array}
\right. 
\nonumber \\
\label{lt20}  \\
X_{-} &=& \frac{ \left | X \right | - X }{ 2 } = 
\left\{
\begin{array}{cc}
0,  & X>0,  \\
\ \\
-X ,  & X<0,  
\end{array}
\right. 
\nonumber 
\end{eqnarray}
\ \\ \noindent
and get

\begin{eqnarray}
\epsilon &=&  
z + y_{+} +  x_{-}  
,
\nonumber \\
\xi^{\prime}  &=&  
z +  x_{+} 
.
\label{lt21} \\
\xi  &=& 
z + y_{-} +  x_{-} 
.
\nonumber 
\end{eqnarray}
\noindent \\
The factors entering Eq.~(\ref{lt17c})  are modified as follows,

\begin{eqnarray}
e^{ - i \epsilon   t_{1}   } 
&\to&
e^{ - i  z   t_{1}   } 
e^{ - i y_{+} t_{1}   } 
e^{ - i   x_{-}  t_{1}   } 
,
\nonumber \\
e^{ i  \epsilon   t_{2}   } 
&\to&
e^{  i  z   t_{2}   } 
e^{  i y_{+}  t_{2}   } 
e^{  i  x_{-}   t_{2}   } 
,
\nonumber \\
e^{ - \frac{\xi^{\prime} }{2 } } 
&\to&
e^{ - \frac{  z }{2 } } 
e^{ - \frac{  x_{+}   }{2 } } 
,
\nonumber \\
e^{  i   \xi^{\prime}   t_{1}   } 
&\to&
e^{  i    z  t_{1}   } 
e^{  i  x_{+}   t_{1}   } 
,
\nonumber \\
e^{ -i \left( \xi^{\prime} \right)^{2}  \frac{\zeta  }{ 4}  } 
&\to&
e^{ -i z^{2}  \frac{\zeta  }{ 4}  } 
e^{ -i z x_{+}   \frac{\zeta  }{ 2}  } 
e^{ -i x_{+}^{2}  \frac{\zeta  }{ 4}  } 
,
\nonumber \\
e^{ i \epsilon \xi^{\prime} \frac{\zeta }{ 2 }  } 
&\to&
e^{ i  z^{2} \frac{\zeta }{ 2 }  } 
e^{ i z y_{+}  \frac{\zeta }{ 2 }  } 
e^{ i  z  x_{+} \frac{\zeta }{ 2 }  } 
e^{ i  z x_{-}  \frac{\zeta }{ 2 }  } 
e^{ i y_{+}  x_{+} \frac{\zeta }{ 2 }  } 
e^{ i  x_{+} x_{-}  \frac{\zeta }{ 2 }  } 
,
\nonumber \\
e^{ - \frac{\xi }{2 } } 
&\to&
e^{ - \frac{  z }{2 } } 
e^{ -  \frac{ y_{-}  }{ 2 } } 
e^{ -  \frac{  x_{-}  }{ 2 } } 
,
\nonumber \\
e^{ -i  \xi   t_{2}   } 
&\to&
e^{ -i   z   t_{2}   } 
e^{ -i y_{-}   t_{2}   } 
e^{ -i   x_{-}  t_{2}   } 
,
\nonumber \\
e^{ i \xi^{2}  \frac{\zeta  }{ 4}  } 
&\to&
e^{ i  z^{2}  \frac{\zeta  }{ 4}  } 
e^{ i   z y_{-}  \frac{\zeta  }{ 2}  } 
e^{ i   z  x_{-}  \frac{\zeta  }{ 2}  } 
e^{ i  y_{-}^{2}  \frac{\zeta  }{ 4}  } 
e^{ i   y_{-} x_{-}  \frac{\zeta  }{ 2}  } 
e^{ i  x_{-}^{2}  \frac{\zeta  }{ 4}  } 
,
\nonumber \\
e^{ - i \epsilon \xi \frac{\zeta }{ 2 }  } 
&\to&
e^{ - i  z^{2} \frac{\zeta }{ 2 }  } 
e^{ - i   z y_{+}  \frac{\zeta }{ 2 }  } 
e^{ - i  z  y_{-} \frac{\zeta }{ 2 }  } 
e^{ - i  z  x_{-}  \zeta  } 
e^{ - i  y_{+} y_{-} \frac{\zeta }{ 2 }  } 
\nonumber \\
&&
e^{ - i   y_{+} x_{-}   \frac{\zeta }{ 2 }  } 
e^{ - i  y_{-} x_{-}   \frac{\zeta }{ 2 }  } 
e^{ - i  x_{-}^{2} \frac{\zeta }{ 2 }  } 
.
\label{lt22} 
\end{eqnarray}
\ \\ \noindent
Substituting Eqs.~(\ref{lt19}) and (\ref{lt22}) into Eq.~(\ref{lt17c}) we obtain,

\begin{eqnarray}
\gamma_{3} &=& 
\int\limits_{ 0 }^{ \infty} d  z 
e^{ -   z  } 
\int\limits_{- \infty}^{ \infty} d  x 
e^{ -  \frac{   |  x  |  }{ 2 } } 
e^{  i  x    t_{1}   } 
e^{ -i x_{+}^{2}  \frac{\zeta  }{ 4}  } 
e^{ - i  x_{-}^{2} \frac{\zeta }{ 4 }  } 
\nonumber \\
&& \times
\int\limits_{- \infty}^{ \infty} dy 
e^{ -  \frac{ y_{-}  }{ 2 } } 
e^{ - i y_{+} t_{1}   } 
e^{  i y  t_{2}   } 
e^{ i  y_{-}^{2}  \frac{\zeta  }{ 4}  } 
e^{ i y_{+} x   \frac{\zeta }{ 2 }  }
. 
\label{lt23} 
\end{eqnarray}
\noindent \\
Here we used $X_{+} X_{-} \equiv 0$. 

The integral over $  z$ is trivial. 
This allows us to represent $ \gamma _{3}$ as a double integral like $ \gamma _{1}$ and $ \gamma _{2}$, that facilitates factorization, 

\begin{eqnarray}
\gamma_{3} &=& 
\int\limits_{- \infty}^{ \infty} d  x 
e^{ -  \frac{   |  x |  }{ 2 } } 
e^{  i  x    t_{1}   } 
e^{ -i x_{+}^{2}  \frac{\zeta  }{ 4}  } 
e^{ - i  x_{-}^{2} \frac{\zeta }{ 4 }  } 
\label{lt24} \\
&& \times
\int\limits_{- \infty}^{ \infty} dy 
e^{ -  \frac{ y_{-}  }{ 2 } } 
e^{ - i y_{+} t_{1}   } 
e^{  i y  t_{2}   } 
e^{ i  y_{-}^{2}  \frac{\zeta  }{ 4}  } 
e^{ i y_{+} x   \frac{\zeta }{ 2 }  }
.
\nonumber 
\end{eqnarray}

\subsubsection{Evaluation of $ \gamma _{2}$}

To rewrite Eq.~(\ref{lt17b}) for $ \gamma_{2}$  we use the following coordinate transformation,

\begin{eqnarray}
\epsilon  &=&  x + y_{+},
\nonumber \\
\xi  &=& x + y_{-} ,
\nonumber \\
\label{lt25} \\
\int\limits_{0}^{ \infty} d \xi 
\int\limits_{0}^{ \infty} d \epsilon 
&\to& 
\int\limits_{0}^{ \infty} dx 
\int\limits_{- \infty}^{ \infty} dy 
.
 \nonumber 
\end{eqnarray}
\noindent \\
The respective exponential factors become the following, 

\begin{eqnarray}
e^{ - i \epsilon  t_{1}   } 
&\to& 
e^{ - i  x t_{1}   } 
e^{ - i  y_{+} t_{1}   } 
,
\nonumber \\
e^{ i \epsilon t_{2}    } 
&\to& 
e^{ i x t_{2}    } 
e^{ i  y_{+} t_{2}    } 
,
\nonumber \\
e^{ - \frac{\xi }{2 } }
&\to& 
e^{ - \frac{ x }{2 } }
e^{ - \frac{ y_{-} }{2 } }
,
\nonumber \\
e^{ -i \xi  t_{2}   } 
&\to& 
e^{ -i x  t_{2}   } 
e^{ -i y_{-}  t_{2}   } 
,
\nonumber \\
e^{ i \xi^{2}  \frac{\zeta  }{ 4}  }  
&\to& 
e^{ i x^{2}  \frac{\zeta  }{ 4}  }  
e^{ i y_{-}^{2}  \frac{\zeta  }{ 4}  }  
e^{ i x y_{-}  \frac{\zeta  }{2}  }  
,
\nonumber \\
e^{ - i \epsilon \xi \frac{\zeta }{ 2 } }
&\to& 
e^{ - i x^{2} \frac{\zeta }{ 2 } } 
e^{ - i x y_{-} \frac{\zeta }{ 2 } } 
e^{ - i x y_{+} \frac{\zeta }{ 2 } } 
e^{ - i y_{+} y_{-}  \frac{\zeta }{ 2 } } 
.
\label{lt26}
\end{eqnarray}
\noindent \\
As a result we get, 

\begin{eqnarray}
\gamma_{2} &=&  - 
\int\limits_{ 0 }^{ \infty } dx
e^{ - \frac{x }{2 } } 
e^{ -i  x t_{1}  }  
e^{- i x^{2}  \frac{\zeta  }{ 4}  }  
\label{lt27} \\
&& \times
\int\limits_{- \infty}^{ \infty} dy 
e^{ -  \frac{ y_{-}   }{ 2 } } 
e^{ - i y_{+}   t_{1}   } 
e^{ i y t_{2}    } 
e^{ i y_{-}^{2}  \frac{\zeta  }{ 4}  }  
e^{ - i x y_{+} \frac{\zeta }{ 2 } } 
.
\nonumber 
\end{eqnarray}
\ \\ \noindent
Here we used that $X_{+} X_{-} \equiv 0$.

\subsubsection{Factorization}

Now we show that the contributions $ \gamma _{1}$ and $ \gamma _{2}$ are canceled by the part of $ \gamma _{3}$ and what remains is a factorizable equation. 
   
Easy to see that $-\gamma_{1}$ is cancelled by the part of $\gamma_{3}$, which arises from the integration area in Eq.~(\ref{lt24}) corresponding to $  x > 0$ and $ y > 0$. 
Indeed, for this area we have, 

\begin{eqnarray}
\gamma_{3}^{\prime} = 
\int\limits_{0}^{ \infty} d  x 
e^{ -  \frac{    x   }{ 2 } } 
e^{  i  x    t_{1}   } 
e^{ -i x^{2}  \frac{\zeta  }{ 4}  } 
\int\limits_{0}^{ \infty} dy 
e^{ - i y t_{1}   } 
e^{  i y  t_{2}   } 
e^{ i y x  \frac{\zeta }{ 2 }  }
.
\label{lt28} 
\end{eqnarray}
\noindent \\
Remember that $X_{-} = 0$ and $X_{+} = X$ for $X>0$. 
The equation above is exactly $ - \gamma_{1}$, see Eq.~(\ref{lt17a}), where we identify $ \xi \sim x$ and $\epsilon \sim y$. 

In addition one can see that $-\gamma_{2}$ is cancelled by the part of $\gamma_{3}$, which arises from the integration area in Eq.~(\ref{lt24}) corresponding to $  x < 0$, where $X _{+} = 0$: 

\begin{eqnarray}
\gamma_{3}^{\prime \prime} &=& 
\int\limits_{ - \infty }^{ 0 } d  x 
e^{ -  \frac{     |x|   }{ 2 } } 
e^{ i  x    t_{1}   } 
e^{ - i  x^{2} \frac{\zeta }{ 4 }  } 
\label{lt29}  \\
&& \times
\int\limits_{- \infty}^{ \infty} dy 
e^{ -  \frac{ y_{-}  }{ 2 } } 
e^{ - i y_{+} t_{1}   } 
e^{  i y  t_{2}   } 
e^{ i  y_{-}^{2}  \frac{\zeta  }{ 4}  } 
e^{ - i x y_{+}  \frac{\zeta }{ 2 }  }
.
\nonumber 
\end{eqnarray}
\noindent \\
Here we change additionally $x \to - x$ and obtain, 

\begin{eqnarray}
\gamma_{3}^{\prime \prime} &=& 
\int\limits_{ 0 }^{ \infty } d  x 
e^{ -  \frac{     x   }{ 2 } } 
e^{ - i  x    t_{1}   } 
e^{ - i  x^{2} \frac{\zeta }{ 4 }  } 
\label{lt30}  \\
&& \times
\int\limits_{- \infty}^{ \infty} dy 
e^{ -  \frac{ y_{-}  }{ 2 } } 
e^{ - i y_{+} t_{1}   } 
e^{  i y  t_{2}   } 
e^{ i  y_{-}^{2}  \frac{\zeta  }{ 4}  } 
e^{ - i x y_{+}  \frac{\zeta }{ 2 }  }
,
\nonumber 
\end{eqnarray}
\noindent \\
which is exactly $ - \gamma_{2}$, see Eq.~(\ref{lt27}). 

Therefore, the sum of all three $ \gamma$'s is given by the part of Eq.~(\ref{lt24}), namely that part, which is defined by the area $  x > 0$ and  $ y < 0$,

\begin{eqnarray}
\sum\limits_{j=1}^{3} \gamma_{j} = 
\int\limits_{- \infty}^{ 0 } dy 
e^{  \frac{ y  }{ 2 } } 
e^{  i y  t_{2}   } 
e^{ i  y^{2}  \frac{\zeta  }{ 4}  } 
\int\limits_{0}^{ \infty} d  x 
e^{ -  \frac{     x    }{ 2 } } 
e^{  i  x    t_{1}   } 
e^{ -i x^{2}  \frac{\zeta  }{ 4}  } 
.
\nonumber \\
\label{lt31} 
\end{eqnarray}
\noindent \\
This equation is represented as the product of two factors, each of which depends on a single time only. 

As the final step we change a sign $  x \to -  x$ in the equation above and substitute it into Eq.~(\ref{lt16}). 
After restoring proper dimensions of time and energy we find that the  correlation function $G ^{(1)} _{cap}$, Eq.~(\ref{lt16}), is cast into the form of Eq.~(\ref{ecf03}) with a wave function given in Eq.~(\ref{ex02}).

\end{document}